\documentclass[CHICAGO,twocolumn]{USG}

\usepackage{anyfontsize}
\usepackage{xcolor}
\usepackage{colortbl}
\usepackage{tikz}
\usetikzlibrary{arrows.meta,positioning,shapes.geometric,angles,quotes,calc}
\graphicspath{{./images/}{./figures/}}

\articletype{RESEARCH ARTICLE}
\journal{}
\volume{0}
\copyyear{2026}
\startpage{1}
\articledoi{}

\usepackage{etoolbox}
\makeatletter
\patchcmd{\@maketitle}{\includegraphics[width=26.5mm]{Wiley_logo.eps}}{}{}%
  {\ClassWarning{USG}{preprint patch: publisher logo not removed}}
\patchcmd{\@maketitle}{\includegraphics[width=56mm]{allergy.eps}}{}{}%
  {\ClassWarning{USG}{preprint patch: sample journal logo not removed}}
\patchcmd{\@@articletype}{\raisebox{-2\@p@t}{\includegraphics[width=0.70in]{images/openaccess}}}{}{}%
  {\ClassWarning{USG}{preprint patch: open-access badge not removed}}
\expandafter\patchcmd\csname\string\@@address\endcsname
  {\setcounter{lastauthor}{4}}{\setcounter{lastauthor}{\number\addcount}}{}%
  {\ClassWarning{USG}{affiliation separator patch not applied}}
\historydates{{\FIfont\strut}}
\def\oddfoot@titlepage@info{%
  \vbox{\hsize\textwidth
    \noindent\rule{\textwidth}{.5\@p@t}\vskip4\@p@t
    {\fontsize{7}{8}\selectfont\hfill{\pagenumfont\thepage\ of \@LastPg}}%
    \vskip10\@p@t}}
\def\evenfoot@titlepage@info{\oddfoot@titlepage@info}
\def\evenhead@headings@info{\head@headings@def{\evenpage@headings@def\hfill}}
\makeatother

\begin{document}

\title{TASCopilot: A provenance-preserving assistant for triple-axis neutron scattering}
\transtitle{TASCopilot: A provenance-preserving assistant for triple-axis neutron scattering}

\author[3]{Carina Jacobson}
\author[2]{Aditya Purohit}
\author[1]{Srikar Kovvali}
\author[1,4]{Lekhnath Poudel}
\author[1,4]{Jeffrey W. Lynn}
\author[1,4,5]{William D. Ratcliff}

\authormark{JACOBSON and PUROHIT \textsc{et al.}}
\titlemark{TAS Co-pilot}

\address[1]{\orgdiv{NIST Center for Neutron Research, }\orgname{National Institute of Standards and Technology, }%
\orgaddress{\city{Gaithersburg, }\state{Maryland 20899, }\country{USA}}}
\address[2]{\orgname{Duke University, }\orgaddress{\state{North Carolina, }\country{USA}}}
\address[3]{\orgdiv{Department of Physics, }\orgname{Purdue University, }\orgaddress{\state{Indiana, }\country{USA}}}
\address[4]{\orgdiv{Maryland Quantum Materials Center, Department of Physics, }\orgname{University of Maryland, }\orgaddress{\city{College Park, }\state{Maryland 20742, }\country{USA}}}
\address[5]{\orgdiv{Department of Materials Science and Engineering, }\orgname{University of Maryland, }\orgaddress{\state{Maryland, }\country{USA}}}

\corres{William D. Ratcliff (\email{william.ratcliff@nist.gov}). Carina Jacobson and Aditya Purohit contributed equally to this study.}
\contributed{Carina Jacobson and Aditya Purohit contributed equally to this study.}

\fundingInfo{Center for High Resolution Neutron Scattering, a partnership between the National Institute of Standards and Technology and the National Science Foundation (Agreement No.~DMR-2010792).}


\abstract[ABSTRACT]{Experimental decisions often depend on evidence scattered across papers, instrument logs, data files, and specialist software. A scientific assistant must connect that evidence without turning a plausible interpretation into an unsupported conclusion. TAS Co-pilot addresses this problem for triple-axis spectroscopy (TAS), which uses neutrons to probe material structure and excitations. It preserves provenance: the sources, processing steps, and review decisions behind each result. Independently maintained scientific tools communicate through structured requests and results; validation rules and human review govern how evidence can support a claim. Language models extract information and explain outcomes but do not approve scientific claims or control the instrument. Dependency-aware reduction executes existing scientific routines and reuses unchanged intermediate results. A retrospective SrMn$_2$Sb$_2$ study connects literature on a related material and historical alignment records to modeling of magnetic excitations in 20 measured scans. A live BT-7 study uses a dedicated instrument interface for fine alignment and 40 adaptively selected measurements. Agreement with a conventional measurement of the same crystal at shared conditions supports the acquisition and analysis chain. These cases demonstrate a methodology for maintaining inspectable evidence across literature, numerical analysis, and measurement at a user facility.}

\maketitle

\section{Introduction}
\label{sec:introduction}

For an experimenter, answering ``Which data support this published figure?'' can require reconciling figure captions, sample identities, instrument configurations, raw-data headers, and calibration records. A relevant scan may have the wrong energy transfer; a detector map may lack the calibration needed to reproduce its axes. A useful scientific assistant must connect these records and explain which claims they support, which candidates they rule out, and what evidence is still missing.

TAS Co-pilot provides this decision-support layer for thermal triple-axis neutron scattering. An experimenter submits a scientific question or a planned measurement through a command-line or browser interface. The system routes it to a bounded workflow: a named procedure with specified inputs and outputs and explicit stop conditions when required evidence is missing. The workflow calls scientific tools and returns an inspectable evidence record. This record preserves provenance: the sources, processing steps, and review decisions behind each result. Language models help extract information and explain the result; scientific calculations, validation rules, and human review determine how that information can be used.

\begin{figure*}[t]
\centering
\begin{tikzpicture}[>=Latex, scale=1.05, transform shape,
  comp/.style={draw, fill=gray!25, font=\scriptsize, align=center},
  blab/.style={font=\scriptsize, inner sep=1.5pt, align=center},
  axlab/.style={font=\scriptsize\itshape, text=gray!45!black},
  beam/.style={->, thick},
  cont/.style={dashed, gray!55, shorten >=0pt}
]
\coordinate (src) at (-1.2,1.8);
\coordinate (M)   at (1.3,0);
\coordinate (S)   at (3.8,1.8);
\coordinate (A)   at (6.3,0);
\coordinate (D)   at (8.8,1.8);
\coordinate (Mc) at ($(src)!1.32!(M)$);
\coordinate (Sc) at ($(M)!1.32!(S)$);
\coordinate (Ac) at ($(S)!1.32!(A)$);

\draw[cont] (M) -- (Mc);
\draw[cont] (S) -- (Sc);
\draw[cont] (A) -- (Ac);

\draw[beam] (src) -- node[blab, above, sloped, pos=0.55]{white beam} (M);
\draw[beam] (M)   -- node[blab, above left]{$\mathbf{k}_i$} (S);
\draw[beam] (S)   -- node[blab, below left]{$\mathbf{k}_f$} (A);
\draw[beam] (A)   -- (D);

\pic[draw, angle radius=0.7cm, angle eccentricity=2.15,
  "{\scriptsize $2\theta_M$ (A2)}"] {angle = Mc--M--S};
\pic[draw=red!65!black, very thick, angle radius=0.7cm, angle eccentricity=2.3,
  "{\scriptsize\color{red!65!black}$2\theta_S = \mathrm{A4}$}"] {angle = A--S--Sc};
\pic[draw, angle radius=0.7cm, angle eccentricity=2.15,
  "{\scriptsize $2\theta_A$ (A6)}"] {angle = Ac--A--D};

\node[comp, minimum width=12mm, minimum height=10mm] (reactor) at (src) {Reactor\\source};
\draw[very thick] ($(M)+(-0.5,0)$) -- ($(M)+(0.5,0)$);
\draw[very thick] ($(A)+(-0.45,0)$) -- ($(A)+(0.45,0)$);
\node[circle, draw, fill=gray!30, minimum size=4.5mm, inner sep=0pt] (sam) at (S) {};
\node[comp, minimum width=9mm, minimum height=6mm] (det) at (D) {Det.};

\node[blab, below=5mm] at (M) {Monochromator\\A1 rotation; sets $E_i$};
\node[blab, above=3mm] at (S) {Sample\\A3 rotation};
\node[blab, below=5mm] at (A) {Analyzer\\A5 rotation; sets $E_f$};
\end{tikzpicture}
\caption{Thermal triple-axis spectrometer in a left--right--left (W) beam configuration. Neutrons from the reactor turn left at the monochromator, which selects the incident energy $E_i$ (and wavevector $\mathbf{k}_i$) by Bragg reflection through the scattering angle $2\theta_M$ (motor A2). The sample scatters through $2\theta_S$, which on NCNR BT-7 is motor \textbf{A4}; the analyzer selects the final energy $E_f$ (wavevector $\mathbf{k}_f$) through $2\theta_A$ (A6) before the detector. A1 and A5 rotate the monochromator and analyzer crystals, respectively, and A3 rotates the sample. These rotations and the scattering angles set the momentum transfer $\mathbf{Q}=\mathbf{k}_i-\mathbf{k}_f$ and energy transfer $E=E_i-E_f$ \citep{lynn2012bt7}. The crystal bars illustrate reflecting planes. Dashed lines mark the undeviated beam direction at each axis.}
\label{fig:triple-axis}
\end{figure*}
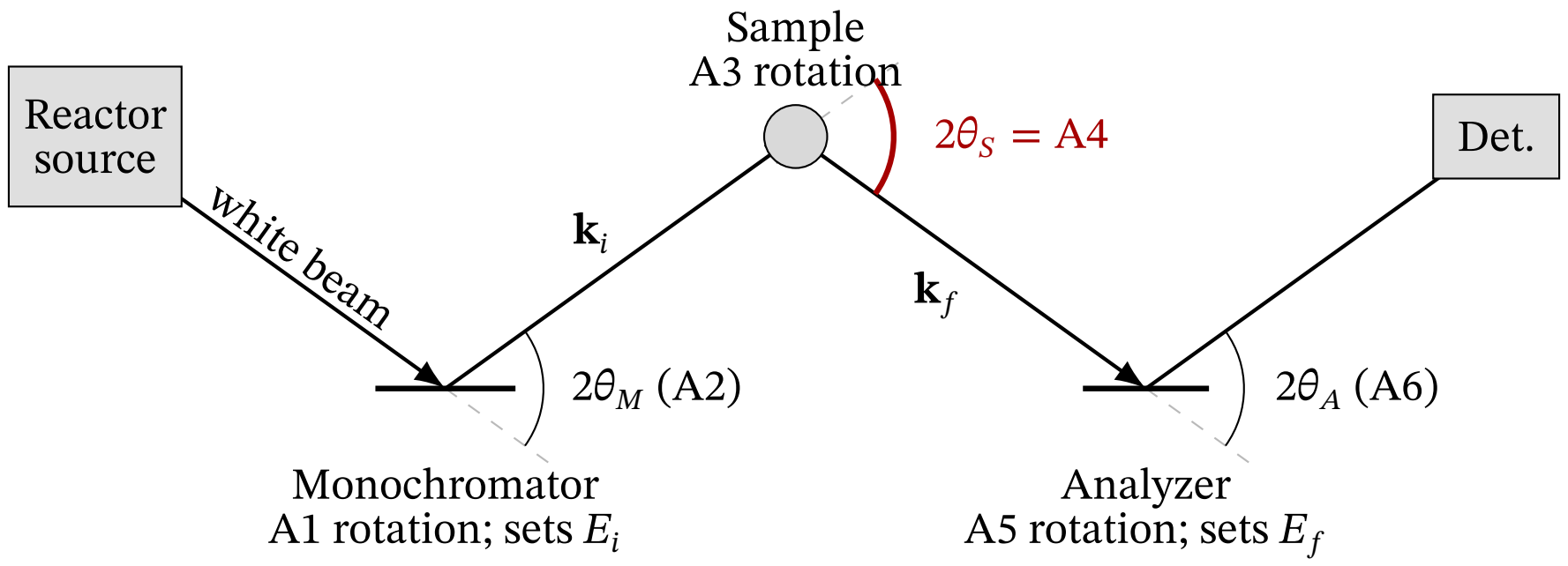

Triple-axis spectroscopy (TAS) measures where neutrons scatter from a sample and how much energy they exchange with it. A monochromator crystal selects the incident neutron energy, the sample scatters the beam, and an analyzer crystal selects the final energy before a detector; the three rotation axes give the technique its name (Figure~\ref{fig:triple-axis}). Each scan records counts as a function of the momentum transfer $\mathbf{Q}=\mathbf{k}_i-\mathbf{k}_f$, where $\mathbf{k}_i$ and $\mathbf{k}_f$ are the incident and final neutron wavevectors, and the energy transfer $E=E_i-E_f$ between neutron and sample. Momentum transfer selects a spatial periodicity and direction in the material. It is expressed in reciprocal space, whose coordinates $(H\,K\,L)$ use the crystal's reciprocal lattice as their basis. Elastic scans ($E=0$) probe static structure through diffraction peaks (Bragg reflections), while inelastic scans ($E\neq0$) probe excitations such as lattice vibrations (phonons) and waves of magnetic-moment motion (spin waves, or magnons). The instrument used here is the BT-7 thermal spectrometer at the NIST Center for Neutron Research (NCNR); its sample scattering-angle motor is A4.

The meaning of a TAS scan depends on its path through reciprocal space and energy transfer, analyzer and detector configuration, collimation, filters, sample environment, and alignment state. That context is distributed across papers, proposals, scan headers, console logs, analysis conventions such as those in DAVE \citep{azuah2009dave}, and operator knowledge. Some settings are manual and may never enter a data file. A publication date alone cannot identify an acquisition session; linking a paper to scans requires instrument records or explicit acquisition information. Connecting these sources requires domain-specific checks and an explicit record of unresolved information.

Related systems address different parts of this problem. MatterChat connects a conversational large language model to atomistic structure representations \citep{tang2026matterchat}. ChemCrow and Coscientist coordinate tools for chemistry and laboratory tasks \citep{bran2024chemcrow,boiko2023coscientist}, building on approaches that interleave model reasoning with external tool calls \citep{yao2023react,schick2023toolformer}. Gaussian-process and active-learning policies steer acquisition at neutron and synchrotron facilities \citep{noack2021gp}, while self-driving laboratories couple planning, execution, and analysis \citep{abolhasani2023selfdriving}. Retrieval-augmented generation (RAG) supplies external evidence to language models \citep{lewis2020rag}, and Graph RAG supplies relational context \citep{edge2024graphrag}. TAS Co-pilot connects these capabilities through records that preserve the source, scientific assumptions, and review status of each result as it passes between tools.

Two cases show what this continuity of evidence provides. A retrospective SrMn$_2$Sb$_2$ analysis connects analog literature and historical alignment records to resolution-aware spin-wave modeling of 20 BT-7 scans, with the processing and model assumptions retained. A live La$_{0.7}$Sr$_{0.3}$MnO$_3$ experiment uses the same contracts to recover a crystal alignment and execute a 40-point adaptive measurement, with a conventional measurement of the same crystal as an independent reference. Human reviews of literature extraction and paper-to-experiment linkage test the quality of the evidence entering these workflows.

The contributions are threefold: a federated architecture that coordinates independently maintained scientific tools through shared contracts and provenance; a semantic layer that connects domain vocabulary, extracted evidence, and human review without merging their authority; and an evaluation spanning known-answer regressions, human labels, retrospective analysis, and live measurement. The implementation centers on NCNR BT-7. Its broader contribution is a way to carry evidence and uncertainty across the software and human decisions that make an experimental result reproducible.

\section{Architecture}
\label{sec:architecture}

\subsection{Design premise}

The architecture separates four roles:

\begin{enumerate}
  \item \textbf{Domain semantics}: the vocabulary of scans, instruments, detectors, reductions, calibrations, materials, and claim states.
  \item \textbf{Evidence}: concrete records from papers, source chunks, data files, scan catalogs, tool runs, derived artifacts, and human review labels.
  \item \textbf{Reasoning workflow}: deterministic or bounded procedures that decide which evidence is relevant, which tools to call, which constraints apply, and which claims remain unsupported.
  \item \textbf{Language generation}: the final transformation of validated evidence and missing-input state into a readable answer.
\end{enumerate}

Language models participate in extraction and in the final explanation. Between those stages, workflows check the extracted records, invoke scientific calculations, and apply review and permission rules. This separation keeps a model's proposed interpretation distinguishable from a validated tool result or a human decision.

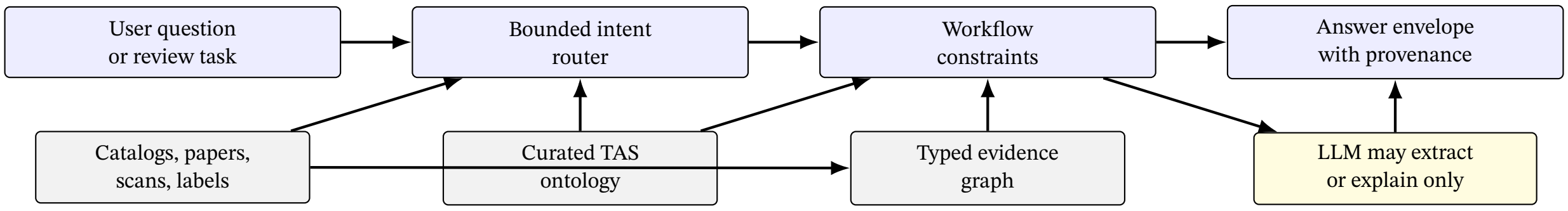
\begin{figure*}[t]
\centering
\begin{tikzpicture}[scale=0.92, transform shape,
  node distance=6mm and 8mm,
  box/.style={draw, rounded corners=1.5pt, align=center, font=\footnotesize, inner sep=4pt, minimum height=8mm},
  layer/.style={box, fill=blue!7, minimum width=38mm},
  evidence/.style={box, fill=gray!10, minimum width=31mm},
  warn/.style={box, fill=yellow!15, minimum width=32mm},
  arrow/.style={-{Latex[length=2mm]}, thick}
]
\node[layer] (question) {User question\\or review task};
\node[layer, right=of question] (router) {Bounded intent\\router};
\node[layer, right=of router] (workflow) {Workflow\\constraints};
\node[layer, right=of workflow] (answer) {Answer envelope\\with provenance};

\node[evidence, below=of router] (ontology) {Curated TAS\\ontology};
\node[evidence, below=of workflow] (graph) {Typed evidence\\graph};
\node[evidence, below=of question] (stores) {Catalogs, papers,\\scans, labels};
\node[warn, below=of answer] (llm) {LLM may extract\\or explain only};

\draw[arrow] (question) -- (router);
\draw[arrow] (router) -- (workflow);
\draw[arrow] (workflow) -- (answer);
\draw[arrow] (stores) -- (router);
\draw[arrow] (stores) -- (graph);
\draw[arrow] (ontology) -- (router);
\draw[arrow] (ontology) -- (workflow);
\draw[arrow] (graph) -- (workflow);
\draw[arrow] (workflow) -- (llm);
\draw[arrow] (llm) -- (answer);
\end{tikzpicture}
\caption{Semantic decision stack. The chat surface is a front door to bounded workflows, not the scientific authority. Ontology, evidence records, validators, and tool outputs constrain the workflow before language generation produces a readable answer.}
\label{fig:semantic-decision-stack}
\end{figure*}

\subsection{From a question to an evidence record}

Consider the opening question about the source of a published figure. For an elastic position-sensitive-detector (PSD) temperature map, the source-triage workflow takes the panel's composition, momentum-transfer range, and temperature range, then searches the experimental catalog. Neutron-diffraction temperature maps of LuFe$_{1-x}$Mn$_x$O$_3$ provide a concrete material context \citep{disseler2015multiferroicity}. Table~\ref{tab:evidence-record} summarizes a synthetic test in which a user-specified manganese fraction of 0.4 conflicts with a panel label of 0.5. It illustrates the returned evidence record for the regression in Section~\ref{sec:evaluation}; it does not identify the published maps' raw source files.

\begin{table*}[t]
\centering
\caption{An illustrative evidence record for the question ``Which scans could support this elastic temperature map?'' The synthetic LuFe$_{1-x}$Mn$_x$O$_3$ example tests composition conflicts and calibration requirements; it is not a reconstruction of published measurements. A useful result includes rejected alternatives and a qualified conclusion, not only a ranked file list.}
\label{tab:evidence-record}
\begin{tabular}{>{\raggedright\arraybackslash}p{0.19\textwidth}>{\raggedright\arraybackslash}p{0.73\textwidth}}
\hline
Record component & Returned information \\
\hline
Unresolved sample identity & The user specifies a manganese fraction of 0.4; the panel label says 0.5. Both interpretations remain active. \\
Candidate evidence & Position-sensitive-detector temperature scans remain candidates; ambiguous material names are not assigned a composition by inference. \\
Rejected alternative & A scan with 3 meV energy transfer cannot source an elastic map, even if its feature occurs at the expected momentum transfer. \\
Missing input & The detector-channel-to-angle calibration lacks review. A map in momentum transfer and temperature cannot yet be reconstructed. \\
Qualified outcome & Candidate source data, not verified source data. Resolve the composition and calibration, then assess reconstruction provenance or obtain human review before making a stronger source claim. \\
\hline
\end{tabular}
\end{table*}

The same record supports a browser report or a conversational explanation. The interface can change without changing the evidence or the conditions under which a stronger claim is allowed.

\subsection{Federated hub-and-spoke structure}

The core maintains typed contracts, tool discovery, permissions, provenance, the curated ontology, evidence records, and workflow orchestration. A typed contract specifies the fields and allowed values of a tool request or result, including its provenance and failure state.

External scientific repositories retain responsibility for data readers, reduction algorithms, structure-factor calculations, resolution models, spin-wave calculations, and adaptive-measurement policies. For example, an alignment-planning workflow obtains nuclear structure factors from GSAS-II through an adapter, combines them with scattering-plane and instrument-reachability constraints, and produces ranked candidate reflections with their supporting inputs and provenance. GSAS-II supplies the crystallographic calculation, while TAS Co-pilot coordinates its use within the planning workflow. This division allows scientific tools to evolve independently while their outputs remain connected through common contracts and traceable handoffs.

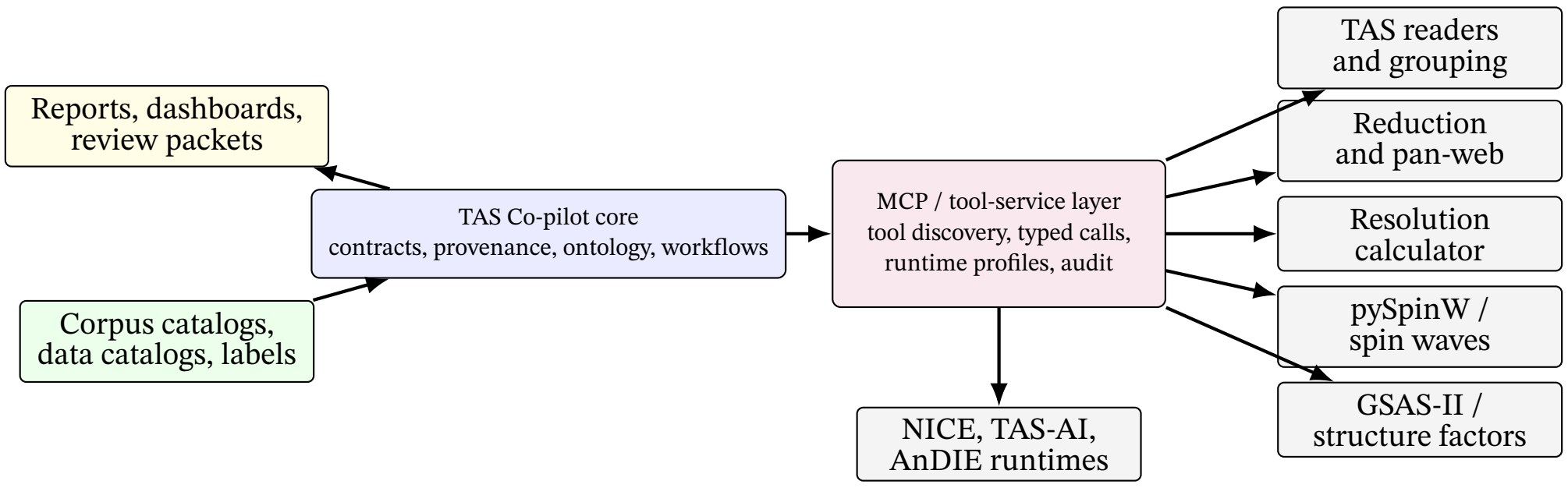
\begin{figure*}[t]
\centering
\begin{tikzpicture}[scale=0.83, transform shape,
  node distance=8mm and 12mm,
  core/.style={draw, rounded corners=2pt, fill=blue!8, align=center, font=\footnotesize, inner sep=5pt, minimum width=37mm, minimum height=9mm},
  mcp/.style={draw, rounded corners=2pt, fill=purple!9, align=center, font=\footnotesize, inner sep=5pt, minimum width=34mm, minimum height=15mm},
  spoke/.style={draw, rounded corners=1.5pt, fill=gray!9, align=center, font=\scriptsize, inner sep=3pt, minimum width=29mm, minimum height=7mm},
  data/.style={draw, rounded corners=1.5pt, fill=green!8, align=center, font=\scriptsize, inner sep=3pt, minimum width=30mm, minimum height=7mm},
  reportbox/.style={draw, rounded corners=1.5pt, fill=yellow!12, align=center, font=\scriptsize, inner sep=3pt, minimum width=33mm, minimum height=7mm},
  arrow/.style={-{Latex[length=2mm]}, thick}
]
\node[core] (core) at (0,0) {TAS Co-pilot core\\contracts, provenance, ontology, workflows};
\node[mcp] (mcp) at (4.6,0) {MCP / tool-service layer\\tool discovery, typed calls,\\runtime profiles, audit};

\node[spoke] (readers) at (8.9,1.9) {TAS readers\\and grouping};
\node[spoke] (reduction) at (8.9,0.95) {Reduction\\and pan-web};
\node[spoke] (resolution) at (8.9,0) {Resolution\\calculator};
\node[spoke] (spinw) at (8.9,-0.95) {pySpinW /\\spin waves};
\node[spoke] (gsas) at (8.9,-1.9) {GSAS-II /\\structure factors};
\node[spoke] (nice) at (4.6,-2.15) {NICE, TAS-AI,\\AnDIE runtimes};
\node[data] (catalogs) at (-3.9,-1.1) {Corpus catalogs,\\data catalogs, labels};
\node[reportbox] (reports) at (-3.9,1.1) {Reports, dashboards,\\review packets};

\draw[arrow] (core) -- (mcp);
\draw[arrow] (mcp) -- (readers);
\draw[arrow] (mcp) -- (reduction);
\draw[arrow] (mcp) -- (resolution);
\draw[arrow] (mcp) -- (spinw);
\draw[arrow] (mcp) -- (gsas);
\draw[arrow] (mcp) -- (nice);
\draw[arrow] (catalogs) -- (core);
\draw[arrow] (core) -- (reports);
\end{tikzpicture}
\caption{Federated hub-and-spoke architecture with an MCP-compatible tool-service layer. The core owns contracts, provenance, ontology, workflow composition, and evaluation. Scientific internals remain in external repositories or runtime profiles. Tool discovery, invocation, runtime selection, audit capture, and typed result normalization happen through the MCP/tool boundary rather than by importing scientific internals into the core.}
\label{fig:federated-architecture}
\end{figure*}

The contracts are also exposed through a Model Context Protocol (MCP) gateway. It validates structured arguments, dispatches requests to the appropriate runtime, and returns results or explicit missing-input records with provenance and audit information. Command-line clients, browser workbenches, and local language models use the same contracts. A resolution calculation therefore retains the same inputs and result schema whether a user requests it interactively or a workflow invokes it during a fit.

\begin{figure*}[t]
\centering
\begin{tikzpicture}[scale=0.86, transform shape,
  node distance=7mm and 10mm,
  layer/.style={draw, rounded corners=2pt, fill=blue!7, align=center, font=\footnotesize, inner sep=5pt, minimum width=38mm, minimum height=9mm},
  mcp/.style={draw, rounded corners=2pt, fill=purple!10, align=center, font=\footnotesize, inner sep=5pt, minimum width=44mm, minimum height=12mm},
  tool/.style={draw, rounded corners=1.5pt, fill=gray!9, align=center, font=\scriptsize, inner sep=3pt, minimum width=29mm, minimum height=7mm},
  record/.style={draw, rounded corners=1.5pt, fill=green!8, align=center, font=\scriptsize, inner sep=3pt, minimum width=33mm, minimum height=7mm},
  arrow/.style={-{Latex[length=2mm]}, thick}
]
\node[mcp, minimum width=54mm] (mcp) {MCP / tool-service layer\\capability inventory, schema validation,\\runtime selection, provenance capture};
\node[layer, above=7mm of mcp] (workflow) {Workflow orchestrator\\selects capabilities, not implementations};
\node[layer, above=6mm of workflow] (surface) {Browser, CLI, bounded copilot router};

\node[tool, below=10mm of mcp] (physics) {Resolution, pySpinW,\\GSAS-II, spurions};
\node[tool] (readers) at (-5.2,0 |- physics) {Readers and\\scan grouping};
\node[tool] (adaptive) at (5.2,0 |- physics) {NICE simulation,\\TAS-AI, AnDIE};

\node[record, left=10mm of mcp] (contracts) {ToolSpec, ToolCall,\\ToolResult contracts};
\node[record, right=10mm of mcp] (runtime) {Runtime profiles:\\conda, container, TACC, service};
\node[record, minimum width=54mm, below=10mm of physics] (audit) {Audit, lineage, review labels,\\derived-artifact provenance};

\draw[arrow] (surface.south) -- (workflow.north);
\draw[arrow] ([xshift=-5mm]workflow.south) -- ([xshift=-5mm]mcp.north);
\draw[arrow] ([xshift=5mm]mcp.north) -- ([xshift=5mm]workflow.south);
\draw[arrow] (mcp.south west) -- (readers.north);
\draw[arrow] (mcp.south) -- (physics.north);
\draw[arrow] (mcp.south east) -- (adaptive.north);
\draw[arrow] (contracts.east) -- (mcp.west);
\draw[arrow] (runtime.west) -- (mcp.east);
\draw[arrow] ([xshift=-23mm]mcp.south) -- ([xshift=-23mm]audit.north);
\draw[arrow] (readers.south) -- (audit.west);
\draw[arrow] (physics.south) -- (audit.north);
\draw[arrow] (adaptive.south) -- (audit.east);
\end{tikzpicture}
\caption{The MCP layer is the uniform service boundary above heterogeneous scientific tools. Workflows ask for capabilities and receive typed results; the layer handles capability discovery, schema validation, runtime dispatch, and provenance/audit capture. This allows a browser, CLI, or future bounded copilot surface to call the same scientific tools without collapsing them into one Python environment or one monolithic codebase.}
\label{fig:mcp-tool-layer}
\end{figure*}
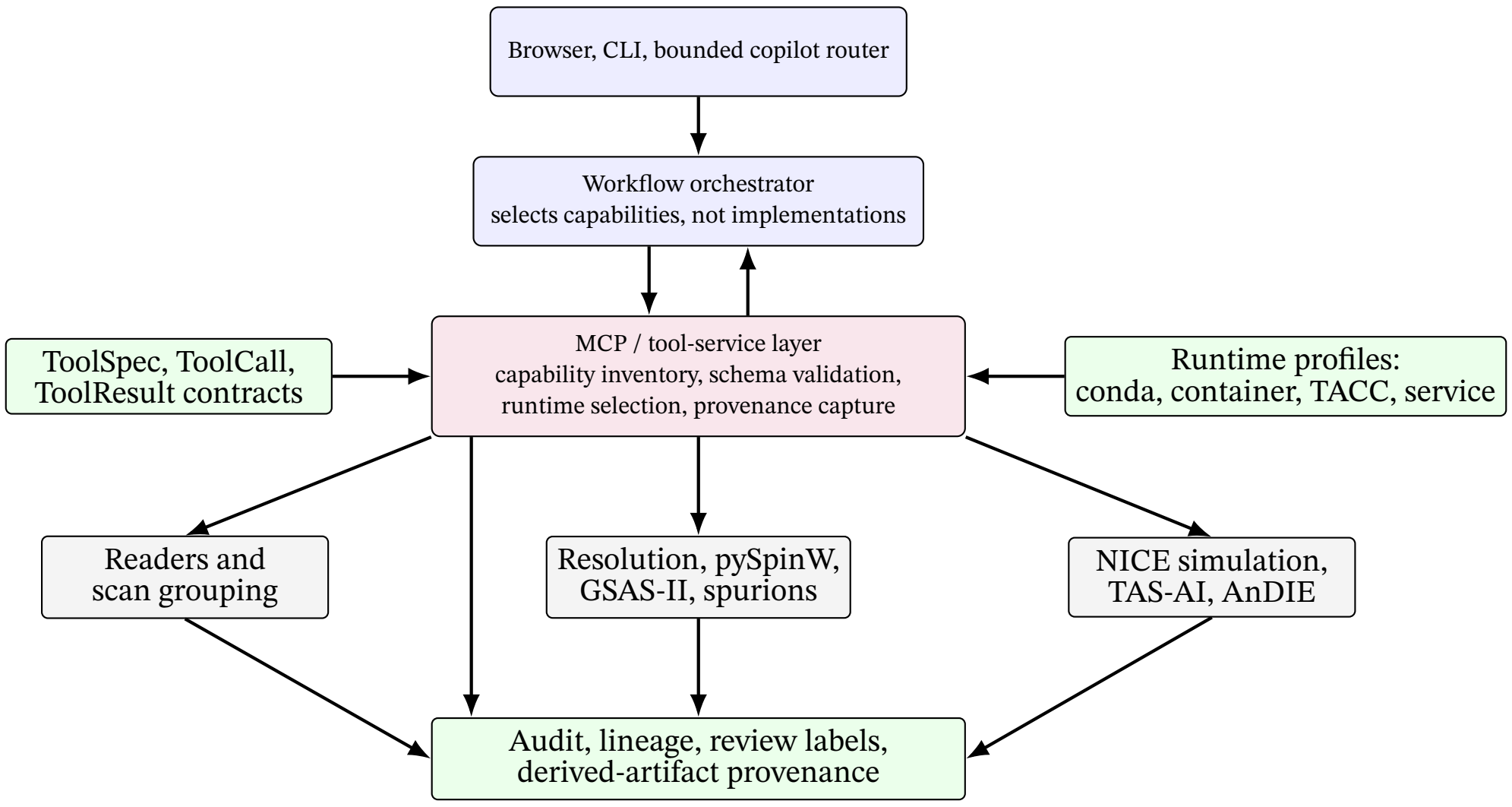

\subsection{Graph RAG, agentic RAG, and MCP tool use}

Standard RAG places retrieved text or records in a model's context \citep{lewis2020rag}; Graph RAG adds relational structure \citep{edge2024graphrag}. TAS Co-pilot retrieves typed one-hop paths connecting a paper to a reported fact, a fact to its supporting passage, or an analog fact to the experimental case it may inform. Each path carries its review status. Agentic RAG allows a model to select retrieval tools and inspect intermediate results before answering \citep{singh2025agenticrag}.

Two bounded mechanisms are used here. The literature-extraction loop retrieves chunks, extracts typed facts, validates their citations, and retrieves again for fact types still missing, subject to a fixed follow-up budget. Separately, Gemma-family local models (Gemma~3 and Gemma~4) \citep{gemma2024,gemma2025v3,gemma2026v4} can read MCP capability descriptions and request curated tools with structured arguments. A model can thus request corpus search followed by source triage, but it cannot accept a scientific claim or reach instrument control through that interface. Extraction quality is measured in Section~\ref{sec:evaluation}; the case studies exercise graph retrieval and tool use without establishing their general accuracy.

\subsection{Ontology, evidence graph, and semantic layer}

The curated TAS ontology gives stable names to operational concepts such as constant-Q scans, fixed-final-energy mode, PSD detection, monitor normalization, and source-support status. Following the vocabulary role described by \citet{gruber1995ontology}, it also assigns retrieval policies: a concept may contribute to ranking, act only as a filter, or be excluded from broad matching. These policies prevent generic concepts from becoming super-nodes that connect nearly every record.

Bulk magnetic properties, including magnetization, anisotropy, exchange stiffness, and coercivity, are drawn from the EMMO-based Magnetic Materials Ontology of the MaMMoS project \citep{hortschitz2026magmo}. A pinned, licensed copy supplies a local concept family with the original identifiers retained. Neutron-specific concepts such as propagation vector, ordered moment per ion, Heisenberg exchange, spin gap, and dispersion remain local. Related quantities retain their distinct units and conventions through explicit mappings. The Supporting Information lists the concept inventory.

Three graph-shaped structures appear in this paper and are not the same thing: the \emph{evidence graph} of this subsection (papers, text chunks, extracted facts, and their review-gated edges), the \emph{reduction DAG} of Section~\ref{sec:workflows} (raw inputs, corrections, operations, and cached artifacts), and the \emph{alignment event record} of Section~\ref{sec:workflows} (a time-ordered log of alignment events read by a state machine). Unqualified, ``the graph'' means the evidence graph. The evidence graph is different from the ontology. The ontology supplies names and policies; the graph connects actual records. For example, it can connect a paper to a text chunk, a chunk to a claim, a claim to a candidate experiment, an experiment to a scan group, a scan group to a data file, a data file to a tool run, and a tool run to a derived artifact.

Edges have types, statuses, confidence labels, evidence, provenance, and optional human review. The graph is therefore not a store of final truth. It is a store of evidence and claim state. A paper chunk may support an instrument claim; a scan header may contradict a figure-source hypothesis; a human label may reject a paper-experiment candidate. The graph preserves all of these as typed records.

Edges enter the graph through two review-gated routes, both now implemented. Deterministic ingest projects catalog records and validated extractions into typed edges (paper-to-chunk, fact-to-supporting-chunk, and analog-prior edges toward a named experimental case). A bounded proposer additionally pairs same-type facts across papers deterministically and asks a local model for one relation judgment per pair from a fixed list of allowed values (consistent, conflicting, same quantity under different conditions, unrelated, or ambiguous); a judgment outside that list is rejected mechanically. Every proposed edge is created with proposed status, and human review flips it to accepted or rejected with reviewer identity and timestamp. Retrieval over the graph folds this review state into ranking: accepted edges carry full weight, proposed edges are down-weighted with raised uncertainty, and rejected edges sink. The same status-weighted graph channel (graph retrieval whose ranking uses the review status) serves the search workflows, the chat surface, and the MCP tool layer, so accepted knowledge and unreviewed candidates remain distinguishable everywhere they appear.

The semantic layer maps user intent to ontology concepts and workflow constraints, constructs retrieval plans, and compiles reports. Validators enforce the domain rules illustrated above: elastic/inelastic compatibility, calibration requirements, composition ambiguity, and review status. Graph retrieval supplies evidence for those checks.

\subsection{Typed extraction rather than free-form graph construction}

Paper and PDF text are treated as untrusted input. Models extract into fixed JavaScript Object Notation (JSON) schemas covering document identity, materials and samples, instrument and sample-environment setup, measurement intent, figures, paper-experiment links, reduction and correction claims, Hamiltonians, and artifact risks. Document text cannot introduce new workflow instructions or arbitrary graph types.

Each record carries source spans, missing fields, validator failures, extraction method, and status. A source span records where the evidence came from in the paper or document. A validator failure is a mechanical reason that a record cannot yet be used, such as a missing instrument name, missing page span, unsupported unit, or unreviewed calibration. This structure allows extracted facts to be reviewed, rejected, or used as low-weight evidence without pretending that the model's confidence is calibrated. Human labels then become the route to empirical calibration.

Citation checks distinguish passages located in the cited chunk, passages found elsewhere in the paper, approximate matches, and unlocated text. The fact normalizer stores located source text and character offsets; the relation proposer uses citation tiers to limit the confidence attached to its judgments. These are mechanical checks of source attribution. A located quotation may still be assigned to the wrong field or material, so its support for the extracted value requires a separate judgment. Section~\ref{sec:gold-set} measures both kinds of error against human labels.

\begin{figure*}[t]
\centering
\begin{tikzpicture}[scale=0.88, transform shape,
  node distance=8mm and 7mm,
  box/.style={draw, rounded corners=1.5pt, align=center, font=\footnotesize, inner sep=4pt, minimum width=26mm, minimum height=8mm},
  state/.style={box, fill=blue!7},
  fail/.style={box, fill=red!8},
  review/.style={box, fill=yellow!14},
  arrow/.style={-{Latex[length=2mm]}, thick}
]
\node[box] (input) {Untrusted input\\paper, header, log};
\node[state, right=of input] (extract) {Typed extraction\\with spans};
\node[state, right=of extract] (validate) {Mechanical\\validators};
\node[review, right=of validate] (proposed) {Proposed\\claim};
\node[review, right=of proposed] (reviewed) {Human review or\\reconstruction};
\node[state, right=of reviewed] (accepted) {Accepted / rejected\\superseded};

\node[fail, below=of validate] (missing) {Missing fields,\\validator failures};
\node[fail, below=of proposed] (ambig) {Ambiguous or\\low-confidence evidence};
\node[state, below=of reviewed] (labels) {Review labels\\calibrate priors};

\draw[arrow] (input.east) -- (extract.west);
\draw[arrow] (extract.east) -- (validate.west);
\draw[arrow] (validate.east) -- (proposed.west);
\draw[arrow] (proposed.east) -- (reviewed.west);
\draw[arrow] (reviewed.east) -- (accepted.west);
\draw[arrow] (validate.south) -- (missing.north);
\draw[arrow] (proposed.south) -- (ambig.north);
\draw[arrow] (reviewed.south) -- (labels.north);
\coordinate (feedback) at ($(ambig.east)!0.5!(labels.west)$);
\coordinate (return) at ([xshift=8mm,yshift=-4mm]proposed.south);
\draw[arrow] (labels.west) -- (feedback |- labels.west)
  -- (feedback |- return) -- (return) -- ([xshift=8mm]proposed.south);
\end{tikzpicture}
\caption{Claim lifecycle. Extracted facts do not become scientific claims directly. They pass through typed validation, preserve missing fields and ambiguity, remain proposed until reviewed or reconstructed, and feed later probabilistic ranking through explicit labels.}
\label{fig:claim-lifecycle}
\end{figure*}
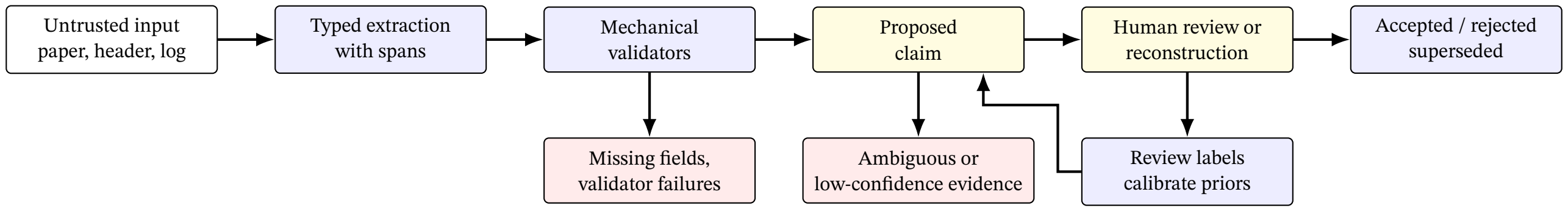

\subsection{Permission and safety boundary}

Conversational interfaces cannot operate the instrument; such requests follow a tested blocked-intent path. The live BT-7 adapter of Section~\ref{sec:case-live} uses one executor, a host named explicitly on the command line, and the control software's trajectory and scan commands. Hardware and software limits remain under instrument control. The co-pilot reads those limits and never writes limits, zeros, or the orientation matrix. An independent kinematic calculation checks each point before counting. An axis policy can exclude motors from optimization and records the reason for each skipped stage. A stop condition ends the chain and produces a record for review; planning output alone does not authorize operation.

Tool contracts declare side effects so that write- or control-capable operations cannot be dispatched as read-only. Corpus text, PDFs, and retrieved snippets are treated as untrusted data and are never executed as code or permitted to override system instructions. New adaptive procedures remain subject to simulation and explicit authorization before live use. These controls define the permitted execution path independently of the language-model interface.
\subsection{Runtime boundaries}

The system assumes that scientific tools have incompatible runtimes. NICE, the NCNR's instrument-control software, may require a Java/container runtime; pySpinW and GSAS-II may require separate Python environments; TAS-AI and AnDIE (a separately developed autonomous neutron-diffraction explorer, a measurement policy like TAS-AI but for diffraction) may run on CPU or GPU clusters; document extraction may involve GROBID \citep{lopez2009grobid}, a parser for scholarly-PDF structure, Docling \citep{auer2024docling}, a layout-aware document-conversion tool, or local model services. TAS Co-pilot therefore treats runtimes as profiles rather than forcing one giant conda environment. The core exchanges typed JSON with adapters and records the software, environment, and tool versions used for each result.

\subsection{Implementation status}

Table~\ref{tab:implementation-status} distinguishes routine facility use from components evaluated in specific workflows. The live experiment-summary page is rebuilt from BT-7 data every 20 minutes; planning pages, catalog workflows, and review tools are also in use. The reduction DAG executes numerical scan reduction, while separate templates record fit lineage and detector previews. The literature-to-graph workflow has been exercised on the two-paper analog case using Gemma on the Vista system at the Texas Advanced Computing Center, and MCP tool use has been exercised at NIST and on Vista. Human verdicts cover a larger 33-paper extraction set, but relation-edge verdicts remain limited to the two-paper pilot. Section~\ref{sec:evaluation} reports these measured results separately from component availability.

\begin{table*}[t]
\caption{Maturity of the main TAS Co-pilot components. \emph{Routine}: used regularly on real BT-7 data. \emph{Implemented}: exercised in specified workflows or evaluations. \emph{Bounded}: operation subject to explicit authorization and constraints.}
\label{tab:implementation-status}
\centering
\footnotesize
\renewcommand{\arraystretch}{1.25}
\begin{tabular}{p{0.27\textwidth} p{0.15\textwidth} p{0.50\textwidth}}
\hline
Component & Status & Notes \\
\hline
Typed contracts, registry, permissions & Routine & Tool/catalog/chunk/edge/label/report contracts; fail-closed permission enforcement. \\
Provenance and audit records & Routine & Per-result software/environment/tool-version capture; audit logs. \\
Curated TAS/BT-7 ontology & Implemented & 130 local operational concepts with retrieval policies (Supporting Information) plus 95 magnetic-materials concepts generated from MagMO. \\
Data and corpus catalogs, evidence graph & Implemented & BT-7 data records; corpus and raw-data catalogs kept separate; review-gated edge proposal and label application. \\
Bounded, ontology-guided retrieval & Implemented & Keyword, ontology, typed graph-traversal (review-status weighted), and fixture channels with super-node diagnostics. \\
Bounded extraction loop (agentic RAG) & Implemented & Retrieve, extract typed facts, validate citations, then re-retrieve for fact types still missing under a fixed follow-up budget; cross-paper relation judgments from a fixed list of values with tiered citation checking; two-paper pilot labels and a 33-paper gold set with human verdicts (Section~\ref{sec:gold-set}). \\
Paper-to-experiment linkage and review store & Implemented & Probabilistic ranking with human-review labels feeding label-informed re-ranking. \\
Figure-source triage (LuFe/Mn/O) & Implemented & Known-answer regression with calibration and elastic/inelastic gates. \\
Experiment dashboards and summaries & Routine & Browser dashboards, overnight and morning summaries, and a live experiment page (order-parameter series, meshes, scan browser) rebuilt every 20 minutes during running experiments. \\
Order-parameter triage and fitting & Implemented & Candidate search plus one-/split-peak model comparison (AIC/BIC). \\
Alignment planning (structure factors) & Implemented & GSAS-II nuclear $F^2$; public dashboard; in-memory CIF handling. \\
Alignment observation (state machine) & Implemented & Epochs and state estimates from historical BT-7 scans. \\
Spurion checks and candidate ranking & Implemented & Deterministic kinematic checks (Section~\ref{sec:workflows}) through a typed boundary when lattice, orientation, and configuration are explicit; candidate features ranked for review by predicted spurion conditions, width proxy, and filter and velocity-selector state. \\
Reduction workflows & Implemented & Executable scan reduction, selective reuse of intermediate results, and replay; separate fit and PSD lineage templates. \\
Resolution and spin-wave adapters & Implemented & External \texttt{rescalculator} and \texttt{tas-pyspinw-adapter} via typed boundaries. \\
NICE control adapter & Implemented & Simulator replay and live one-point trajectories and \texttt{findPeak} chains on BT-7; a real host must be named explicitly, every point passes an independent $|\mathbf{Q}|$ pre-flight check, and alignment stages end in recorded stop conditions. \\
MCP gateway over the bounded tools & Implemented & Flat-argument tools, structured missing inputs, provenance capture; direct evidence-graph search with review-status filtering. \\
Local-model (Gemma) MCP tool use & Implemented & Gemma drives the MCP tools; exercised at NIST and on Vista (TACC) with preliminary route-quality and prompt-injection evaluations. \\
Live-instrument operation & Bounded & One executor, named host, control-software commands only; limits, zeros, and orientation are never written; chat surfaces remain blocked. \\
\hline
\end{tabular}
\end{table*}

\section{Scientific workflows}
\label{sec:workflows}

The workflows follow an experimenter's tasks: find relevant data, establish their relationship to a scientific question, inspect and reduce scans, plan an alignment, and evaluate or execute a measurement plan. Each produces machine-readable records and a readable report linked to its inputs and tool calls. The cases in Section~\ref{sec:case-studies} show how these records connect across tasks.

\subsection{Catalog and retrieval workflows}

Data catalogs hold file metadata, scan trajectories, detector modes, scan groups, and experiment summaries. Separate corpus catalogs hold papers, source passages, figures, panels, and extracted claims. Retrieval combines keyword, ontology-guided, and graph-based searches; diagnostics flag concepts that match too many records. The data-catalog layer is designed to interoperate with NeXus, the standard data format for neutron, x-ray, and muon experiments \citep{koennecke2015nexus}, and materials-data exchange patterns such as OPTIMADE \citep{andersen2021optimade}, while retaining metadata peculiar to historical TAS files.

The separation between corpus and data catalogs is essential. A paper chunk may say that a measurement used BT-7, a PSD detector, or a fixed final energy. A data-file catalog may say that a directory contains scans with a particular \texttt{ExptID}, detector mode, scan family, title, and timestamp. A source-data claim is only formed when a workflow connects those two evidence streams through typed candidate records. This prevents a retrieval hit in a PDF from silently becoming an accepted experimental fact.

\subsection{Paper-to-experiment and source-data linkage}
\label{sec:case-linkage}

Paper-to-experiment linkage ranks candidates using material identity, instrument, date priors, experimenters, titles, source passages, scan families, and human-review labels. The linker can rank scan groups within an experiment as well as whole experiments, because one identifier may contain several samples, alignment scans, calibrations, and follow-up measurements. Unreviewed scores express ranking evidence and are not calibrated probabilities.

The output remains a proposed \texttt{PaperExperimentLink}. Review dashboards expose the paper, candidate experiment metadata, scan summaries, and reviewer fields. Accepted and rejected labels are then ingested into a unified review store. A negative human label is treated as ground truth for that reviewed row; positive labels raise confidence but still preserve the supporting evidence channels and review provenance. Section~\ref{sec:gold-set} reports the reviewed linkage results.

\begin{figure*}[t]
\centering
\begin{tikzpicture}[scale=0.9, transform shape,
  node distance=10mm and 8mm,
  box/.style={draw, rounded corners=1.5pt, align=center, font=\footnotesize, inner sep=4pt, minimum width=23mm, minimum height=8mm},
  store/.style={box, fill=gray!10},
  decision/.style={draw, diamond, aspect=2.2, align=center, font=\footnotesize, inner sep=1pt},
  arrow/.style={-{Latex[length=2mm]}, thick}
]
\node[store] (paper) {Paper corpus\\chunks};
\node[box, right=of paper] (panel) {Figure/panel\\metadata};
\node[box, right=of panel] (query) {Compile\\source query};
\node[store, below=of query] (data) {TAS data\\catalog};
\node[box, right=of query] (triage) {Source triage\\ranking};
\node[decision, right=of triage] (review) {Reviewed\\evidence?};
\node[box, right=12mm of review] (accepted) {Accepted or\\candidate link};
\node[box, below=of triage] (reject) {Rejected candidates\\with reasons};
\node[box, below=of review] (labels) {Review labels\\and priors};
\draw[arrow] (paper.east) -- (panel.west);
\draw[arrow] (panel.east) -- (query.west);
\draw[arrow] (query.east) -- (triage.west);
\draw[arrow] (data.north) -- (triage.south west);
\draw[arrow] (triage.east) -- (review.west);
\draw[arrow] (review.east) -- node[above,font=\scriptsize]{yes/no} (accepted.west);
\draw[arrow] (triage.south) -- (reject.north);
\draw[arrow] (review.south) -- (labels.north);
\coordinate (feedback) at ($(reject.east)!0.5!(labels.west)$);
\coordinate (return) at ([xshift=8mm,yshift=-5mm]triage.south);
\draw[arrow] (labels.west) -- (feedback |- labels.west)
  -- (feedback |- return) -- (return) -- ([xshift=8mm]triage.south);
\end{tikzpicture}
\caption{Paper-to-experiment and figure-source linkage are evidence-preserving ranking workflows. Corpus chunks, panel metadata, and TAS data catalogs remain separate inputs. The workflow produces proposed links, rejected candidates, and review-label records; accepted source-data claims require reconstruction/provenance or human review.}
\label{fig:paper-source-linkage}
\end{figure*}
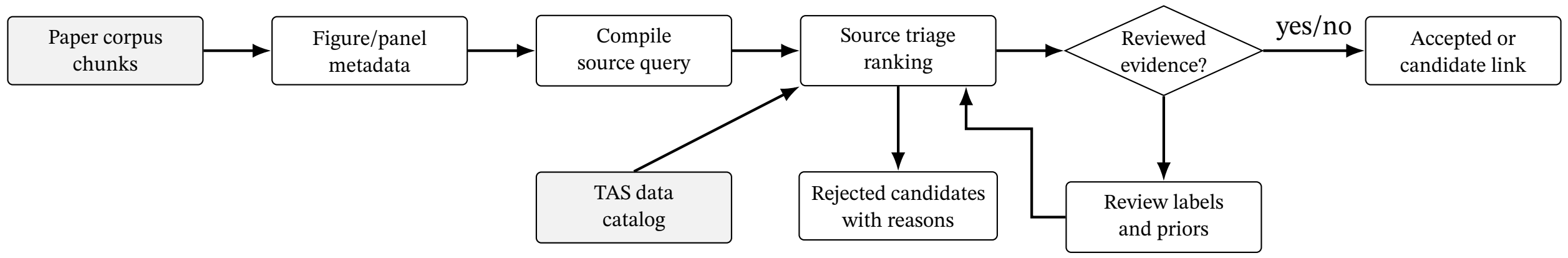

Figure-source linkage adds panel-specific constraints: which scans or reconstruction artifacts could have produced this panel? Captions and panel labels supply composition, geometry, and measurement conditions for the search in Figure~\ref{fig:paper-source-linkage}. Conflicts between the user's sample description and the panel label remain separate hypotheses. The LuFe/Mn/O example introduced in Section~\ref{sec:architecture} tests this behavior alongside the elastic-scan and calibration requirements.

\subsection{Experiment dashboards and overnight summaries}

Browser dashboards bring scan plots, metadata, and review labels together. They summarize scan families, candidate order parameters, inelastic series, possible spurions, and alignment evidence from the workflow outputs. Daily and experiment-level reports retain the same source-file lineage, allowing a scan identified in a summary to be traced into a later analysis. Review labels enter the shared store used by the linkage and alignment workflows.

\subsection{Panweb peak analysis and order-parameter workflows}

Panweb (\url{https://tripleaxis.org/panweb/}) provides the interactive pattern used here: load temperature-dependent scans, inspect the data, compare single-peak and split-peak models, and examine the resulting intensity trend. TAS Co-pilot classifies the scan family, normalizes by monitor, preserves alternative background models, and records the fits as candidate order-parameter or structural-splitting evidence. Alignment and slit-scan analysis use the same pattern of model comparison and review, with models appropriate to their different physical questions.

\subsection{Alignment assistance}
\label{sec:case-alignment}

At \url{https://tripleaxis.org/alignment/} and in the local workbench, an experimenter selects a structure and scattering plane, views reciprocal-space geometry and common background materials, and inspects candidate alignment reflections. The same planning calculation is available from the command line.

The planning side uses local structure information and a separate GSAS-II runtime \citep{toby2013gsasii} to calculate neutron nuclear structure factors (the structure factor $F$ of a reflection fixes its intensity through $|F|^2$); TAS Co-pilot does not implement its own crystallographic structure-factor engine. Candidate reflections are filtered by scattering-plane membership, sample two-theta reachability, relative $F^2$, and out-of-plane angle computed with the reciprocal metric tensor. Background and mounting materials such as aluminum and copper are treated as scoring evidence rather than rejection rules. The goal is to help the experimenter choose robust peaks and recognize possible powder-ring, substrate, mount, or $\lambda/2$ ambiguities.

Historical scan headers and console logs provide a second source of alignment evidence. Search moves, peak searches, tilt and translation scans, and software-state changes become a time-ordered event record. A state machine reconstructs alignment epochs from that record: a remount or zero reset can invalidate an earlier assignment without deleting the observation that supported it. The resulting state estimates and review packets can be compared with the planned reflections. The same event schema records the live alignment chain in Section~\ref{sec:case-live}.

Fine-alignment analysis compares one-, two-, and three-peak Gaussian models using the Bayesian information criterion (BIC), flags excessive peak multiplicity for review, and preserves uncertainty about twins, broad mosaic, powder background, and out-of-plane projections. Model selection requires an explicit fitting backend; without one, the workflow returns a lightweight preview. Slit scans use ramp/saturation and S-curve models because their question is sample centering rather than Bragg peak shape. These outputs are planning evidence and do not approve motor motion.

\subsection{Reduction DAGs and correction state}

Reduction choices can change as an experiment is interpreted: a different background scan or revised normalization may require parts of an analysis to be repeated. Following the graph-and-cache approach of Reductus \citep{maranville2018reductus}, TAS Co-pilot represents these dependencies as a directed acyclic graph (DAG). Raw inputs remain unchanged, and corrections and parameters are recorded separately. The graph both describes the numerical reduction and determines its execution.

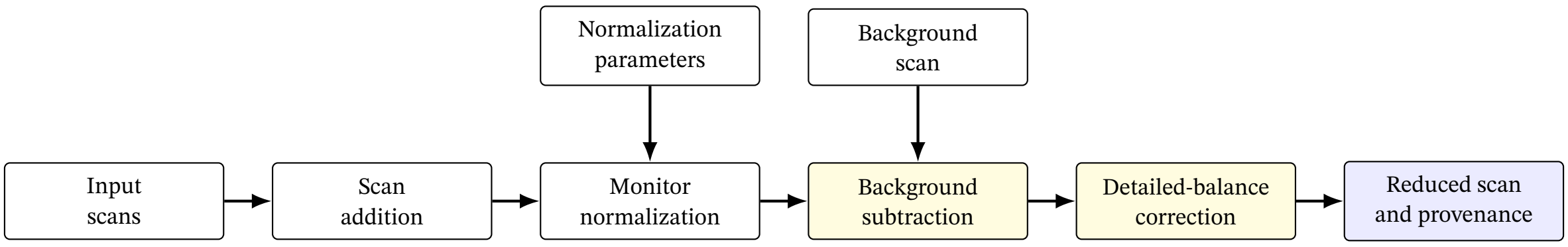
\begin{figure*}[t]
\centering
\begin{tikzpicture}[scale=0.95, transform shape,
  node distance=8mm and 5mm,
  box/.style={draw, rounded corners=1.5pt, align=center, font=\footnotesize, inner sep=4pt, minimum width=23mm, minimum height=8mm},
  affected/.style={box, fill=yellow!15},
  cache/.style={box, fill=blue!8},
  arrow/.style={-{Latex[length=2mm]}, thick}
]
\node[box] (raw) {Input\\scans};
\node[box, right=of raw] (combine) {Scan\\addition};
\node[box, right=of combine] (norm) {Monitor\\normalization};
\node[affected, right=of norm] (subtract) {Background\\subtraction};
\node[affected, right=of subtract] (balance) {Detailed-balance\\correction};
\node[cache, right=of balance] (result) {Reduced scan\\and provenance};
\node[box, above=of norm] (parameters) {Normalization\\parameters};
\node[box, above=of subtract] (background) {Background\\scan};
\draw[arrow] (raw) -- (combine);
\draw[arrow] (combine) -- (norm);
\draw[arrow] (parameters) -- (norm);
\draw[arrow] (norm) -- (subtract);
\draw[arrow] (background) -- (subtract);
\draw[arrow] (subtract) -- (balance);
\draw[arrow] (balance) -- (result);
\end{tikzpicture}
\caption{An executable scan-reduction graph using existing scientific routines. Revising the background scan recomputes subtraction and the subsequent correction (shaded), while allowing the combined and normalized scans to be reused. The resulting scan retains its propagated uncertainties and provenance. The sequence illustrates dependency tracking, not a universal prescription for correction order.}
\label{fig:reduction-dag}
\end{figure*}

Numerical nodes invoke existing external routines for reading scans, combining them, normalizing to monitor counts, subtracting backgrounds, and applying detailed-balance corrections. Each result becomes an input to its dependent calculations, retaining the uncertainties propagated by the scientific routines and the provenance of the tool calls. Saved intermediate results are reused when their inputs, correction records, and software versions are unchanged; a revision recomputes the affected dependent steps (Figure~\ref{fig:reduction-dag}). A failed or unavailable calculation prevents its dependents from proceeding, and replay remains subject to the same permission checks as the original execution.

Separate scan-family templates preserve the selected order-parameter fit branch and provide detector-channel by temperature previews. They record analysis choices without replacing the fitting or detector-reconstruction workflows; conversion of PSD channels to $Q$ requires a reviewed channel-to-A4 calibration.

\subsection{Adaptive-measurement interfaces}
\label{sec:case-adaptive}

TAS-AI exists as a separate active-learning project for autonomous spin-wave spectroscopy \citep{ratcliff2026tasai}. TAS Co-pilot does not absorb that implementation. Instead it provides typed contracts that let TAS-AI propose measurement points, run analytic examples, and interact with NICE, in simulation or live, through a bounded adapter. In this division of labor, TAS-AI owns the measurement policy, NICE owns the instrument API and its limits, and TAS Co-pilot records the plan, evidence, provenance, and safety state and executes each point with a pre-flight check (Section~\ref{sec:case-live}).

Spin-wave and resolution calculations follow the same division of responsibility. An adapter invokes the external Python implementation of SpinW-style calculations \citep{toth2015spinw}; a separate resolution calculator supplies broadening and convolution calculations and is also available at \url{http://drneutron.org/res/}. The SrMn$_2$Sb$_2$ case combines these tools in a joint intensity fit, recording the model, resolution scenario, software versions, and fitted endpoint at each handoff.

Spurion checks follow the same pattern. A spurion is a spurious peak that is not scattering from the sample at the nominal momentum and energy transfer: a Bragg reflection reached by a higher harmonic of the monochromator or analyzer wavelength, a powder line from aluminum, copper, or molybdenum in the sample mount and environment, or Bragg scattering of the incident or final beam by the sample combined with incoherent scattering elsewhere in the spectrometer (the Currat--Axe spurion). Each has a kinematic signature, so for a given lattice, scattering plane, fixed-energy configuration, and scan trajectory the conditions under which each can appear are computed and compared with the features in the data. The checker is a Python port of the computational core of the Spurion2 module of DAVE \citep{azuah2009dave}, the NCNR data-analysis suite, behind a \texttt{tas-spurion} boundary; the dashboard ranks candidate features for review by the proximity of these predicted conditions together with width, filter, and velocity-selector evidence.

\section{Case studies}
\label{sec:case-studies}

Two cases test the architecture across a complete sequence of scientific tasks.
The retrospective SrMn$_2$Sb$_2$ case follows evidence from analog literature
and historical instrument records into a resolution-aware fit of measured
scans. The live La$_{0.7}$Sr$_{0.3}$MnO$_3$ case tests whether the same
contracts and records support alignment and adaptive measurement against
independent experimental references. The first tests continuity of evidence
through analysis; the second tests it through execution.

\subsection{An integrated real-data walkthrough:
\texorpdfstring{SrMn$_2$Sb$_2$}{SrMn2Sb2}}
\label{sec:case-srmn2sb2}

We used a retrospective SrMn$_2$Sb$_2$ experiment as an end-to-end test because
the analysis requires evidence that normally remains separated across papers,
crystallographic files, instrument logs, raw scans, resolution metadata, and
specialized modeling codes. Measurements used the BT-7 spectrometer
\citep{lynn2012bt7}, and the direct experimental evidence came from experiment
\texttt{24858}, measured in two sessions in 2018. The sample was a 0.5~g single
crystal with a measured antiferromagnetic N\'eel temperature of 114.6(6)~K
(see Figure~1 in Supporting Information~S1). The current analysis was
retrospective, performed after the 2018 experiment; the
live-instrument case is Section~\ref{sec:case-live}. The scientific task is to
infer interactions between magnetic moments from the measured spin-wave
spectrum. A Hamiltonian specifies these interactions; the spectrum tests its
form and parameter values. This case tests how the co-pilot connects source
discovery, model selection, and quantitative fitting while preserving the
evidence and scientific choices behind the analysis.

The workflow began with two CaMn$_2$Sb$_2$ papers supplied by the user outside
the managed corpus. Powder diffraction established candidate magnetic-
structure information, while inelastic scattering supplied an analogous
honeycomb spin-wave Hamiltonian \citep{bridges2009camn2sb2,mcnally2015camn2sb2}.
The private copies remained outside the repository. The bounded extraction
loop of Section~\ref{sec:architecture} (retrieve, extract, validate, then
re-retrieve for what is still missing) converted them in an isolated runtime
and proposed 43 typed facts spanning magnetic structure, propagation vector, spin
direction, ordered moment, and exchange interaction. These records retained
source spans and review status. They were treated as reviewable \emph{analog}
evidence for choosing model terms, never as direct facts about SrMn$_2$Sb$_2$.
In a preliminary human pass, 11 of the 43 fact candidates were marked correct,
and five of seven proposed cross-paper relation edges were rejected. These
small-pilot results explain why candidate records never entered the evidence
graph as accepted truth; results on a larger gold set are reported in
Section~\ref{sec:gold-set}.

The same case connected planning to the observed instrument history. Given the
SrMn$_2$Sb$_2$ structure and scattering plane, the alignment workflow ranked
strong $(0\,0\,2)$-type reflections and $(1\,1\,0)$ using the external GSAS-II
structure-factor boundary. Historical BT-7 headers and console logs then
provided an independent record of what actually happened: alignment on
$(0\,0\,2)$, checks at $(0\,0\,4)$ and $(1\,1\,0)$, a July remount, plane
flattening, and repeated recentering. A temporal state machine reconstructed
those epochs from rocking, theta--two-theta, tilt, translation, and software-
state events. It also produced a fine-alignment replay plan, run against the simulator
because the experiment had ended, with explicit human-review and stop points; the replay was not an instruction
to the instrument.

At larger time scales, the same catalog generated eleven retrospective daily
summaries, an experiment-level scientific narrative, and session-separated
dispersion maps. The dispersion report assembled 10,353 inelastic points into
28 reciprocal-space/energy panels without pooling the two measurement
sessions. This supplied a concrete test of shared lineage: an individual scan
could appear in an overnight summary, an alignment-state reconstruction, an
experiment timeline, and a later model fit while retaining the same source
record and distinct derived-artifact provenance.
Table~\ref{tab:srmn2sb2-provenance} summarizes the evidence class and retained
authority boundary at each handoff.

\begin{table*}[t]
\centering
\caption{Evidence and decision boundaries in the SrMn$_2$Sb$_2$ walkthrough.
Each phase produces a typed artifact that can become input to the next phase,
but its evidence class and human-review status are preserved.}
\label{tab:srmn2sb2-provenance}
\begin{tabular}{>{\raggedright\arraybackslash}p{0.13\textwidth}p{0.22\textwidth}p{0.25\textwidth}p{0.22\textwidth}}
\hline
Phase & Evidence used & Derived product & Boundary retained \\
\hline
Analog literature & Two user-supplied CaMn$_2$Sb$_2$ papers & 43 typed candidate facts & Analog prior; human review required \\
Alignment planning & Sr structure and scattering plane & Ranked reflections and scan ranges & Retrospective planning only \\
Alignment replay & BT-7 scans and console state & Epochs, state estimates, replay plan & Observed history kept separate from proposals \\
Experiment summary & Cataloged scans and sessions & 11 daily reports; experiment narrative & Source-file lineage retained \\
Dispersion mapping & Inelastic scan points & 28 panels; 10,353 points & Sessions not pooled \\
Hamiltonian & Reviewed analog model family and Sr topology & pySpinW model and position-derived starts & External scientific-code authority \\
Intensity fit & 20 scans, backgrounds, resolution samples & Exchange estimates and sensitivity comparisons & Model and normalization assumptions retained \\
\hline
\end{tabular}
\end{table*}

For quantitative modeling, a human selected from the analog literature a model
family with nearest-neighbor and next-nearest-neighbor in-plane exchange,
interlayer exchange, and easy-plane anisotropy on the explicit
SrMn$_2$Sb$_2$ topology. TAS Co-pilot did
not reimplement the scientific engines. The external pySpinW runtime supplied
linear-spin-wave energies and magnetic intensities through a SpinW-style model
\citep{toth2015spinw}; the external resolution layer supplied scrambled-Sobol
samples of a Cooper--Nathans four-dimensional triple-axis resolution model
\citep{cooper1967resolution}; and stock Bumps
supplied optimization and sampling. Runtime gates recorded the source revision,
Hamiltonian convention, scan selection, parameter handoffs, and backend.
pySpinW offers an optional Rust-accelerated intensity backend beside its
reference Python one; a regression test that runs the two side by side found
them in disagreement for this model, so the Rust backend was excluded and the
final calculation used the Python intensities (\texttt{use\_rust=False}).

An instrument does not measure an exact point in momentum and energy: its
finite resolution averages a neighborhood around that point. The model must
be averaged over the same neighborhood before comparison with the data. In TAS,
that neighborhood is represented by a resolution ellipsoid; its tilt relative
to a spin-wave branch can change both the apparent width and position of a
peak. The analysis jointly modeled 20 scans
containing 799 retained points after a reviewed quality mask removed one
detector-spike point. Backgrounds were fixed from preceding scan-local peak
fits, each scan had an independent multiplicative scale, and the calculated
intensity included the Mn$^{2+}$ magnetic form factor squared. The calculation used
64 quasi-Monte Carlo resolution samples per measured point, one in-plane spin
orientation, and a documented legacy resolution scenario whose collimation
(beam angular acceptance) and mosaic (spread of crystal orientations) metadata
were incomplete. Bumps supplied the parameter optimization, with the reported
values obtained from the final Levenberg--Marquardt refinement. Supporting
Information S1 records the optimization sequence and sensitivity comparisons.

The joint fit gives a good overall description of the 20 scans, reproducing
the main peak positions and much of the profile structure along
$(H\,H\,2)$, $(H\,H\,3)$, and $(0\,0\,L)$ with a common set of four
Hamiltonian parameters. With 20 independent scan scales, it gives
$\chi^2_\nu=1.76$ for 775 nominal degrees of freedom. In the implemented
exchange convention, the fitted exchange parameters are
$S J_1=10.3$ meV, $J_2/J_1=0.156$
($S J_2=1.61$ meV), and $S J_c=0.37$ meV. Four predeclared combinations
of two starting parameter sets (each derived from the measured peak
positions) and two normalization policies assessed sensitivity to these
analysis choices (Supporting Information S1). All four solutions retain
dominant nearest-neighbor exchange, with weaker next-nearest-neighbor and
interlayer couplings. The anisotropy parameter is more sensitive to these
choices, as detailed in S1.

\begin{figure*}[t]
\centering
\includegraphics[width=\textwidth]{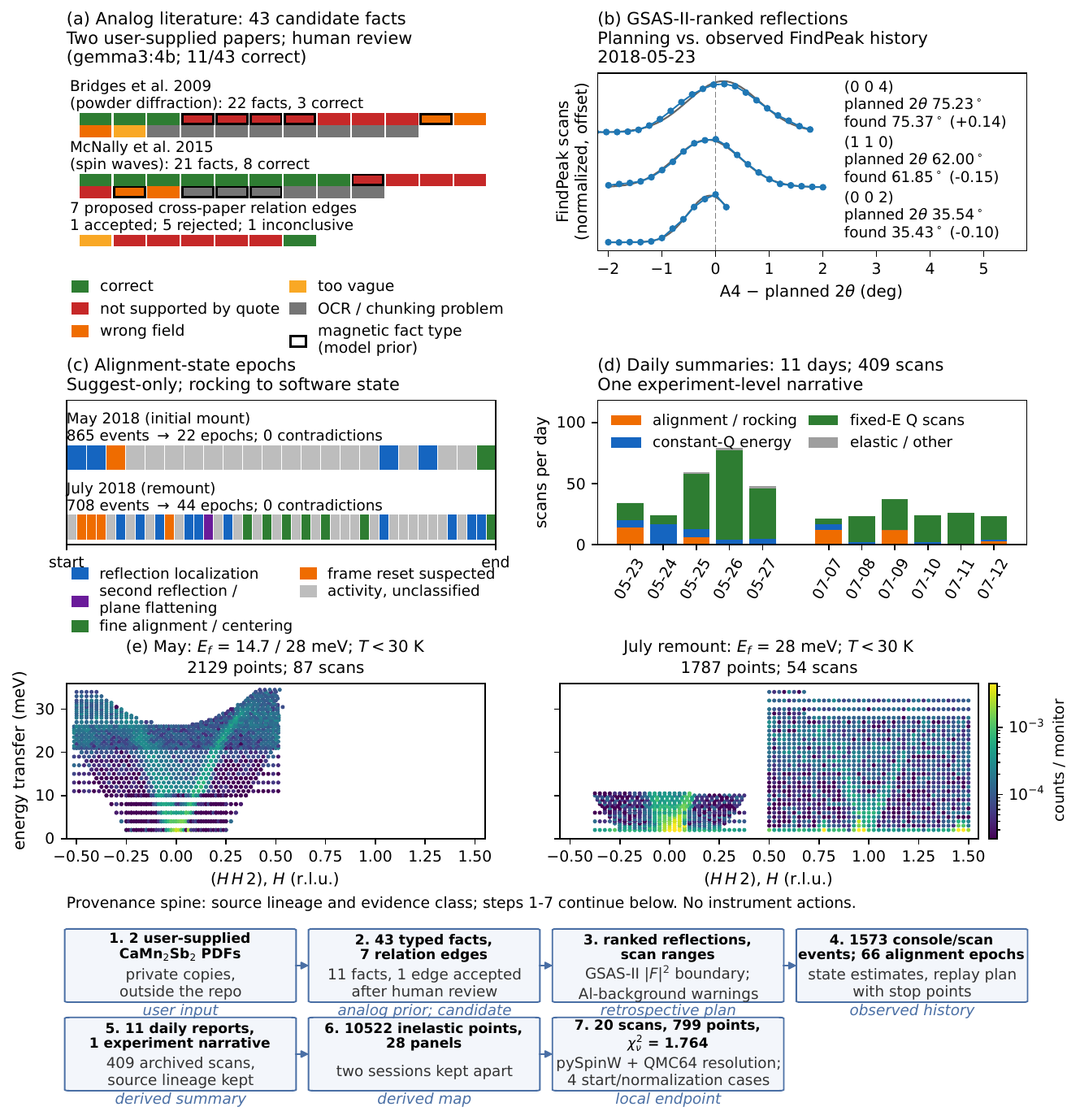}
\caption{A real-data workflow reconstructed across evidence layers for BT-7
experiment \texttt{24858}. (a) Forty-three candidate facts extracted from two
user-supplied CaMn$_2$Sb$_2$ papers, with the human verdicts that kept them at
candidate status (11 correct; the magnetic fact types used as model priors are
outlined). (b) Retrospective alignment planning (GSAS-II-ranked reflections)
against the FindPeak scans actually run on 2018-05-23: the planned $2\theta$ of
$(0\,0\,2)$, $(1\,1\,0)$ and $(0\,0\,4)$ matched the observed centers to within
$0.15^\circ$. (c) Alignment-state epochs reconstructed from console-log and
scan events for the May mount and the July remount; every estimate is
suggest-only and flagged for human review. (d) Retrospective daily summaries
over both sessions. (e) Session-separated dispersion slices along
$(H\,H\,2)$. The provenance spine records the evidence class retained at each
handoff. The replay ran against the simulator; the experiment had ended in 2018, so
no instrument action was involved.}
\label{fig:srmn2sb2-evidence-trace}
\end{figure*}

\begin{figure*}[t]
\centering
\includegraphics[width=\textwidth]{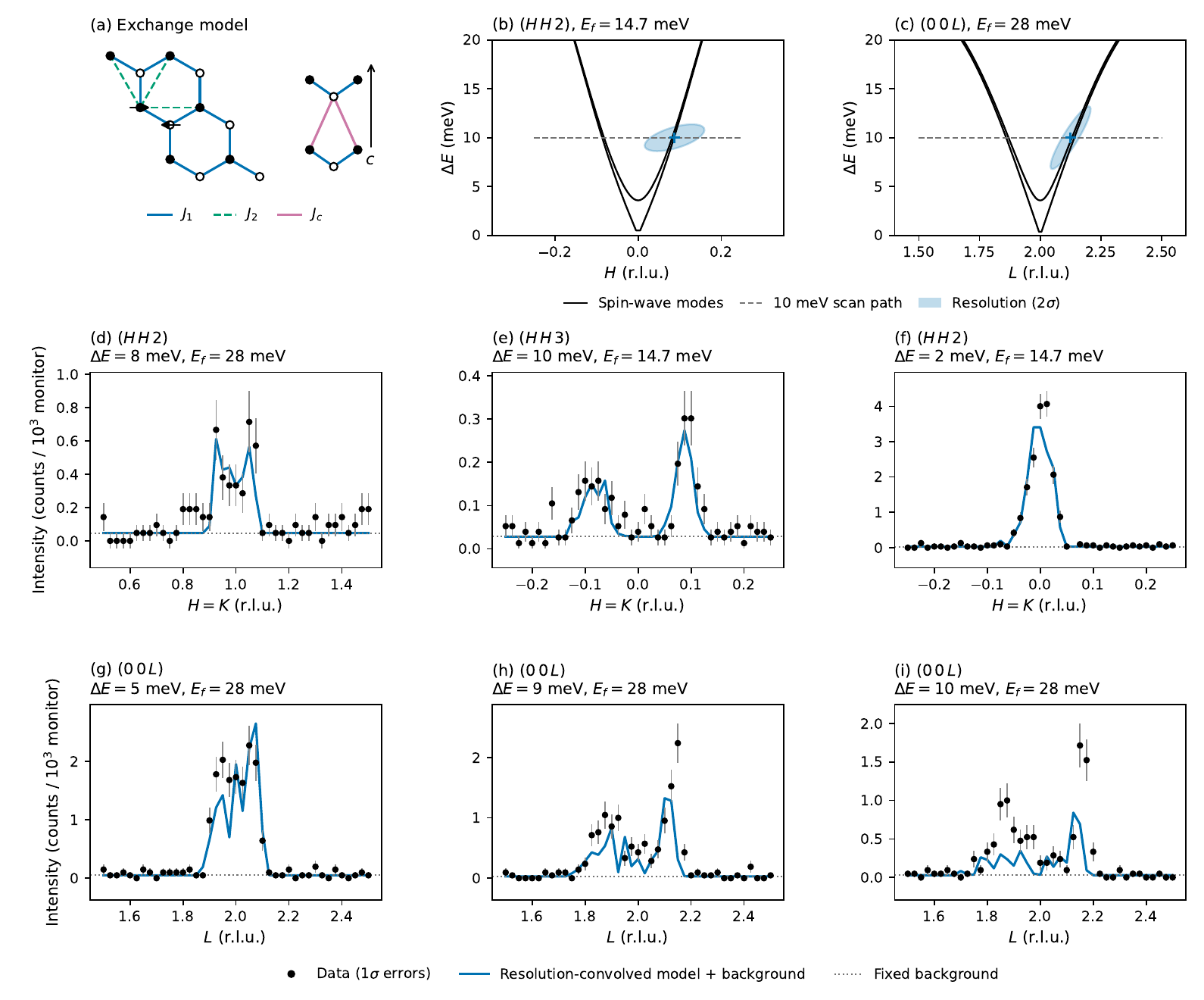}
\caption{Resolution-aware spin-wave modeling of SrMn$_2$Sb$_2$.
(a) Exchange paths in the Mn honeycomb-bilayer model, with trigonal space
group $P\overline{3}m1$ (No.~164): $J_1$ connects the two Mn
sublattices, $J_2$ connects sites within a sublattice, and $J_c$ couples layers
along $c$. Filled and open circles distinguish the sublattices; arrows mark
antiparallel in-plane spins.
(b, c) Low-energy portions of the calculated dispersion along $(H\,H\,2)$
and $(0\,0\,L)$. Dashed lines mark 10 meV scan trajectories. At each marked
scan point, the blue ellipse shows the approximate $2\sigma$ contour of the
projected instrument resolution.
(d--i) Six of the 20 jointly fitted scans (799 points in total). Black points
are measured intensities with one-standard-deviation errors; blue curves are
the resolution-convolved model plus the fixed background (dotted).
The common Hamiltonian reproduces the main peaks and much of the profile
structure, while the 9 and 10 meV
$L$ scans in (h, i) show the clearest remaining line-shape discrepancies.
Each scan has an independent intensity scale; fit parameters and assumptions
are given in the text.}
\label{fig:srmn2sb2-resolution-fit}
\end{figure*}

The clearest remaining line-shape discrepancies occur in the
$(0\,0\,2)$-centered $L$ scans at $\Delta E=9$ and 10 meV with $E_f=28$ meV
(Figure~\ref{fig:srmn2sb2-resolution-fit}h,i). These profiles provide a focused
test for further checks of instrument resolution, domain averaging, and data
reduction. The good overall agreement and the identifiable residual structure
make the fit a useful basis for further analysis of the archived experiment.
Figures~\ref{fig:srmn2sb2-evidence-trace} and
\ref{fig:srmn2sb2-resolution-fit} show the evidence trace and the
resolution-aware model test; both are generated from the public archive files
and the exported fit endpoint by scripts in the manuscript repository. In
these and all subsequent plots of measured scans, error bars represent one
standard deviation.

The case links literature-guided model selection to a quantitative fit of
archived measurements. With independent scan scales, the comparison tests peak
positions and line shapes. The scan selection, processing choices, and fitted
parameters remain connected, allowing the analysis to be revisited and
extended as additional instrument metadata become available.
\subsection{A live case: alignment and adaptive measurement of
\texorpdfstring{La$_{0.7}$Sr$_{0.3}$MnO$_3$}{La0.7Sr0.3MnO3} on BT-7}
\label{sec:case-live}

Cold-neutron measurements established nearly gapless, long-wavelength spin
waves in La$_{0.7}$Sr$_{0.3}$MnO$_3$ at low temperature and a spin-wave
stiffness that decreases with temperature
\citep{vasiliudoloc1998spindynamics}.
The SrMn$_2$Sb$_2$ walkthrough is retrospective. The same architecture was
used prospectively on BT-7 (experiment \texttt{202608/14}, a
La$_{0.7}$Sr$_{0.3}$MnO$_3$ crystal at room temperature, fixed
$E_f=14.7$ meV) through the instrument-control adapter. A conventional
constant-$\mathbf{Q}$ series measured on the same crystal in 2017 provided an
independent reference. TAS Co-pilot coordinated the scientific tools and
instrument software, checked measurement conditions, and retained the results
and their provenance. Two workflows were exercised:
fine alignment of a Bragg reflection, and adaptive point-by-point measurement
in which an external policy (TAS-AI) chooses each point and TAS Co-pilot
executes, guards, and records it. The instrument's own limits were read and
never written, and every motion went through the control software's
trajectory and \texttt{findPeak} commands rather than direct motor writes.
Table~\ref{tab:live-results} summarizes the alignment and measurement checks.

\begin{table}[t]
\caption{Alignment and measurement checks in the live
La$_{0.7}$Sr$_{0.3}$MnO$_3$ case on BT-7.}
\label{tab:live-results}
\centering
\footnotesize
\begin{tabular}{@{}>{\raggedright\arraybackslash}p{0.52\columnwidth}>{\raggedright\arraybackslash}p{0.40\columnwidth}@{}}
\hline
Quantity & Result \\
\hline
Recovered \texttt{a3} against the orientation-matrix prediction & 0.08$^\circ$ (one fifteenth of the 1.13$^\circ$ rocking width) \\
Repeatability of \texttt{a3} over three chains, one aborted mid-sweep & 0.001$^\circ$ \\
Agreement of two independent chains & 0.0005$^\circ$ in \texttt{a3}, 0.0007$^\circ$ in \texttt{a4} \\
Tilt correction found and its effect & 0.5$^\circ$ adjustment; rocking peak 0.8\% higher \\
Adaptive run & 40 of 40 proposed points counted, no refusals, 96 min \\
Intensities against the 2017 measurement at shared $\delta q$ & Coincide with no fitted scale factor \\
Quadratic stiffness coefficient from the live points & $D=135.3$ meV\,\AA$^2$ (fit includes a quartic correction) \\
\hline
\end{tabular}
\end{table}

\begin{figure*}[p]
\centering
\includegraphics[width=\textwidth]{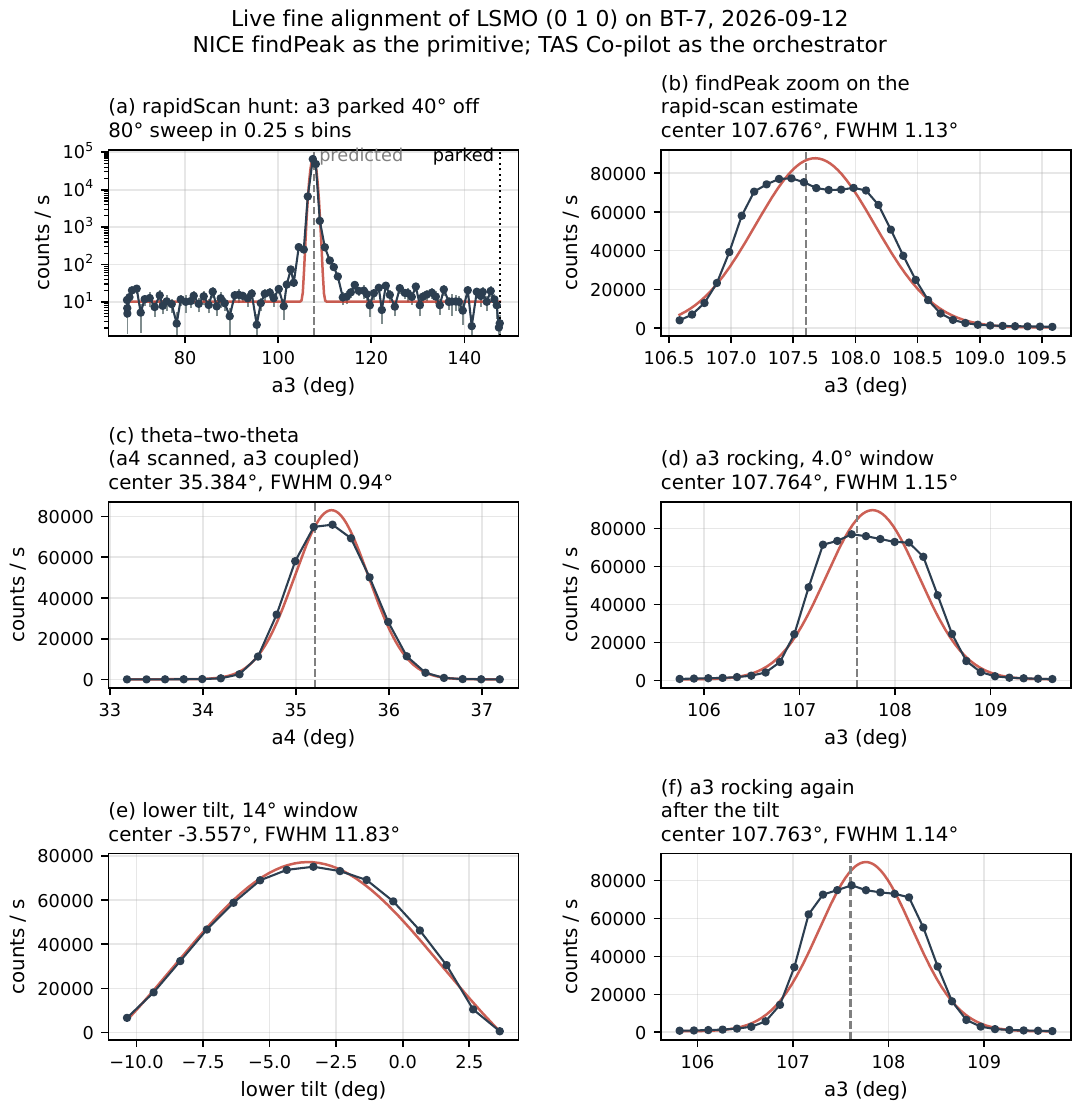}
\caption{Live fine alignment of the La$_{0.7}$Sr$_{0.3}$MnO$_3$ $(0\,1\,0)$
reflection on BT-7 (experiment 202608/14, 2026-09-12), driven remotely by TAS
Co-pilot through the NICE control software. (a) With \texttt{a3} parked
40$^\circ$ from the predicted position (dashed), a continuous \texttt{rapidScan}
locates the peak in 32 s; the bin-integrated counts are position estimates
good to about a bin, not fits. (b) \texttt{findPeak} zoom on the rapid-scan
estimate. (c) Theta--two-theta with \texttt{a3} coupled. (d) \texttt{a3}
rocking in the agreed 4$^\circ$ window. (e) The lower tilt arc, the one
perpendicular to $\mathbf{Q}$ for this reflection, in a 14$^\circ$ window.
(f) \texttt{a3} rocking after the tilt. Dashed lines are the predictions from
the orientation matrix in use; red curves are Gaussian fits drawn for
orientation only, and the flat tops of the rocking profiles at
$7.5\times10^4$ counts/s are detector dead time, which does not shift the
center. The chain recovers \texttt{a3} to 0.08$^\circ$ of the prediction and
repeats to 0.001$^\circ$; data files are public in the NCNR archive
\citep{ncnrdata}.}
\label{fig:lsmo-live-alignment}
\end{figure*}

\noindent\textbf{Live fine alignment.}
Alignment establishes how the crystal's axes sit relative to the beam, so a
requested reciprocal-space position corresponds to the intended part of the
sample's scattering pattern. A rocking scan rotates the sample through a
Bragg peak; a theta--two-theta scan moves the sample and detector together;
tilt scans adjust the scattering plane.
The alignment state machine of Section~\ref{sec:case-alignment} was designed
for fine adjustment from an existing gross alignment. The live test therefore
started from a known orientation, discarded it deliberately, and asked the
co-pilot to recover it: the sample rotation \texttt{a3} was parked 40$^\circ$
away from the predicted $(0\,1\,0)$ position and a continuous \texttt{rapidScan}
swept back across 80$^\circ$ (Figure~\ref{fig:lsmo-live-alignment}a). The
co-pilot then ran the agreed chain, \texttt{findPeak} zoom, theta--two-theta,
\texttt{a3} rocking, the tilt arc perpendicular to $\mathbf{Q}$, and
\texttt{a3} again, with the control software's own \texttt{findPeak} as the
scanning and fitting primitive and TAS Co-pilot supplying the reflection,
the stage order, the windows, an independent centroid cross-check of every
fit, and the stop rules. The recovered $\texttt{a3}$ agreed with the prediction
from the existing orientation matrix to 0.08$^\circ$, one fifteenth of the
1.13$^\circ$ rocking width, and repeated to 0.001$^\circ$ across three chains,
including one whose coarse sweep had been aborted mid-flight. Two independent
chains agreed in \texttt{a3} to 0.0005$^\circ$ and in \texttt{a4} to
0.0007$^\circ$. The tilt moved by 0.5$^\circ$ across a broad intensity profile,
and the final rocking scan peaked 0.8\% higher as a result.

The alignment record captures the checks and constraints governing each stage.
NICE declines to move on a
strong Bragg peak because its reduced $\chi^2$ tolerance is exceeded by counting
statistics, so the co-pilot judges each fit itself (significant excess over
background, width between the step and three quarters of the window, center
away from the window edge, agreement with a background-subtracted centroid) and
moves only on a pass. A window with no peak produces a recorded stop.
Every stage was emitted as a typed event in the same schema the state
machine reads for historical scans, so the live chain was classified by the
same code (reflection localization, then plane flattening) and subjected to the
same prediction and contradiction checks, using the recorded static angles
and unit cell. An
\emph{axis policy} declared the motors that must not be optimized in this run,
each with its reason (pre-sample slit readbacks reset by a controller power
cycle, post-sample slits irrelevant for inelastic work, translations not
reversible from a distance); stages on those axes are skipped and reported
rather than silently omitted. The workflow is designed for fine alignment;
states requiring a new gross alignment or interpretation of multiple crystal
domains are handed to the experimenter.

\noindent\textbf{Live adaptive measurement.}
The target is the magnon dispersion: how the energy of a magnetic excitation
changes with momentum. Unlike a predefined scan grid, an adaptive policy uses
the counts already collected to choose the next momentum--energy point.
The adaptive interface of Section~\ref{sec:case-adaptive}, previously exercised
only against the NICE simulator, ran live for 40 points of the $(0\,K\,0)$
magnon dispersion, 60 s each, in 96 minutes, with 40 of 40 points counted and
no refusals. TAS-AI \citep{ratcliff2026tasai} owned the policy in three
phases: an initial grid to explore the range, a log-Gaussian-process model to
predict intensity from the observations, and a physics-informed phase that
scores disagreement among three ferromagnetic dispersion hypotheses.
TAS Co-pilot owned the executor, which turned each proposed point
into a one-point control-software trajectory, pre-checked the solved angles
against an independent $|\mathbf{Q}|$ calculation before counting, read the
detector back, and appended a trace row with the policy's metadata and the
achieved position. One file per point is not an accident of implementation:
the instrument's $\mathbf{Q}$-to-angle solve was found to run one point
behind on energy transfer within a multi-point trajectory, so a trajectory
whose energy changes is only correct at its first point. The pre-flight guard
is what turned that discovery from lost data into a refused point.

\begin{figure*}[p]
\centering
\includegraphics[width=\textwidth]{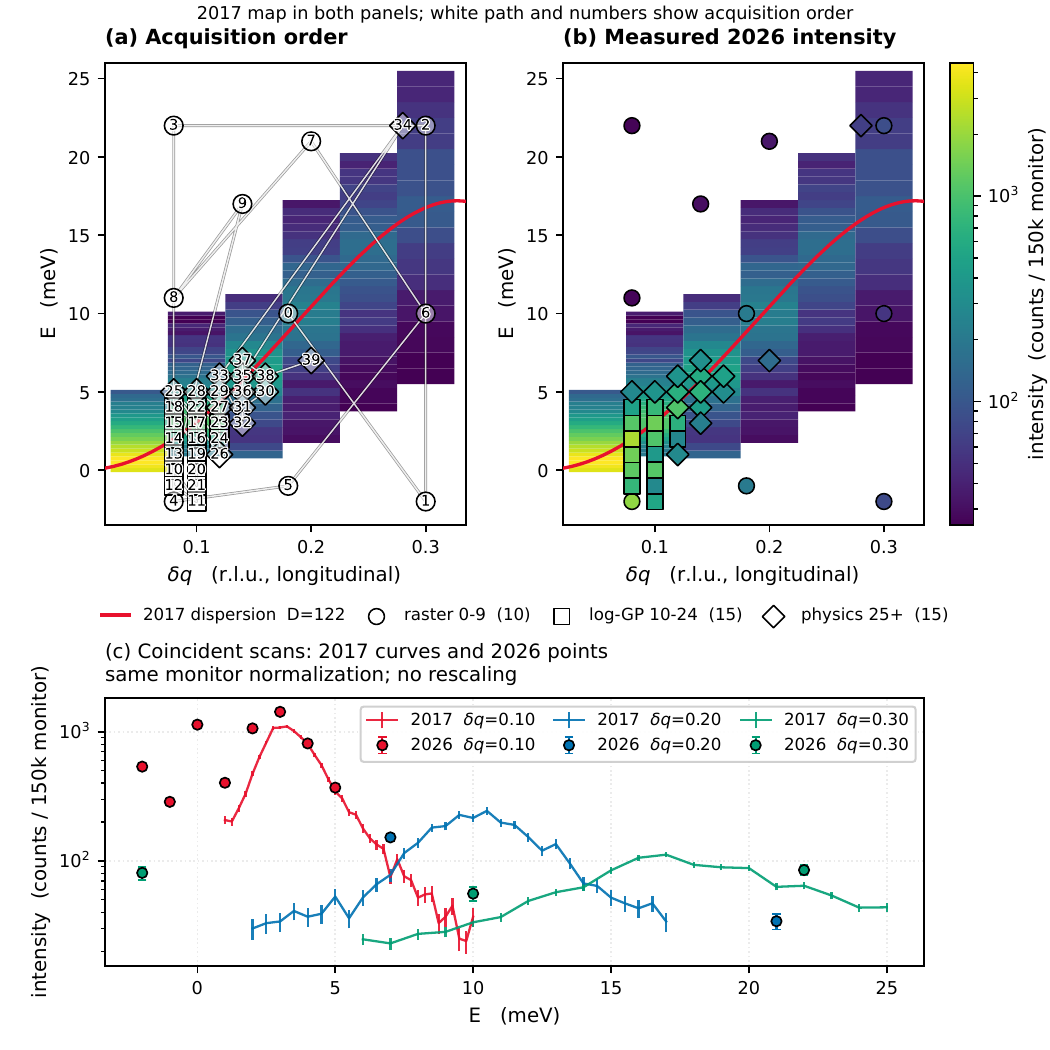}
\caption{The live adaptive run against the conventional 2017 measurement of
the same La$_{0.7}$Sr$_{0.3}$MnO$_3$ crystal at 300 K (BT-7 experiment 23816,
six constant-$\mathbf{Q}$ scans, color map). (a) The 40 live points in
acquisition order, marker shape giving the policy phase (raster, log-GP,
physics), joined by the acquisition path; the red curve is the 2017
dispersion. (b) The same points filled with their own 2026 intensity on the
2017 color scale, so a marker that vanishes into the map agrees with it.
(c) Direct overlay at the three $\delta q$ values the two datasets share,
both in counts per $1.5\times10^5$ monitor with no rescaling. The agreement
supports the combined acquisition and normalization. The acquisition path
records the concentration of adaptive steps at small $\delta q$ alongside
the initial grid's broader coverage.}
\label{fig:lsmo-tasai-vs-2017}
\end{figure*}

Figure~\ref{fig:lsmo-tasai-vs-2017} places the 40 live points on the 2017
conventional map of the same crystal, with counts normalized to incident-beam
monitor readings and no fitted scale factor. Here $\delta q=K-1$ is the
displacement from the $(0\,1\,0)$ reference point, expressed in
reciprocal-lattice units (r.l.u.). Where the two overlap, at $\delta q$ of
0.10, 0.20 and 0.30 r.l.u., the live intensities sit on the 2017 curves. This provides an
independent check of the combined acquisition and reduction chain, including
the normalization and the $K$-versus-$H$ symmetry mapping. It does not isolate
the contribution of each component; the energy-lag guard is tested by the
recorded refusal described above. The dispersion is captured (the intensity
maximum moves from about 2 meV at $K=1.08$ to 5--6 meV at $K=1.14$), and the
quadratic stiffness coefficient is $D=135.3$ meV\,\AA$^2$. This parameter
describes the initial rise of excitation energy with momentum; the fitted
dispersion includes a quartic correction, as specified in the Supporting
Information. The fit omits instrumental-resolution averaging; its parameters
and the sampling coverage are recorded in the Supporting Information.
The point sequence, intensities, and analysis choices remain linked in the
measurement record. Both figures of this case are regenerated from
the public archive files alone by scripts in the repository; the run's trace
is rebuilt from the file names, which carry the step, $\mathbf{Q}$ and $E$ of
every point.

\section{Evaluation}
\label{sec:evaluation}

Known-answer tests assess numerical consistency and enforcement of workflow
constraints; human review assesses extracted facts and proposed
paper-to-experiment links. These evaluations complement the scientific
results in Section~\ref{sec:case-studies}. Retrieval over the evidence graph and the
model's tool selection are described in Section~\ref{sec:architecture} and
exercised in the case studies; they are not measured here.

\subsection{Workflow consistency}

An executable reduction sequence was compared with direct calls to the same external routines for scan addition, monitor normalization, background subtraction, and detailed-balance correction. On synthetic scan inputs, the two paths agreed in counts, propagated uncertainties, monitor values, scan coordinates, and recorded treatments. Revising a background or normalization parameter recomputed the affected calculations while reusing unchanged results. These checks establish numerical consistency with the underlying scientific tools.

Known-answer source-triage checks use synthetic, sanitized records for three
model LuFe$_{1-x}$Mn$_x$O$_3$ panels with $x=0.25$, 0.4, and 0.5
(Table~\ref{tab:evidence-record}). The published material study
\citep{disseler2015multiferroicity} supplies physical context; the regression
tests workflow constraints, not reproduction of its measurements.
The workflow preserves competing composition assignments, rejects inelastic
scans as sources of elastic scattering, and requires reviewed calibration
before reconstruction. Even the strongest surviving match remains a candidate
until reconstruction provenance or human review supports identification.

Missing metadata and unavailable tools produce explicit outcomes rather than inferred replacements. Failed numerical steps block dependent calculations, and reuse of a saved result remains subject to current permissions and tool availability. Conversational requests cannot reach instrument control. These checks enforce evidence and permission boundaries independently of the language model.

\subsection{Human-review labels}
\label{sec:gold-set}

Paper-to-experiment links, extracted facts, source-triage candidates,
order-parameter candidates, spurion candidates, and alignment events are
routed through one review-label store. A negative label is treated as truth
for the reviewed candidate; a positive one raises confidence but keeps the
evidence channels and the review provenance. Two layers have enough labels
to be scored: extraction and linkage.

\paragraph{Extraction: the gold set}
After the two-paper pilot of Section~\ref{sec:case-srmn2sb2}, a larger gold
set was extracted from BT-7-linked papers with a larger local Gemma model
(Gemma~4, 26B parameters). After wrong downloads and duplicate ingests were
identified through title and document-text checks and quarantined, it holds
492 facts from 33 papers. Each paper enters the
catalog as a bounded reading surface rather than as full text: its title, its
abstract when one is detected, one 3000-character window around
instrument-setup, experimental, and data-reduction keywords, and its figure
captions, a median of eight chunks and about 15,000 characters per paper. A
verdict of \emph{not found} therefore means absent from that surface, and the
extractor's recall is bounded by it.

Two of the coauthors reviewed the facts. One labeled a first pass; the other
then reviewed the same facts and continued through the remainder, so that 334
of the 492 facts (68\%), from all 33 papers, carry a final verdict with one of
five values (correct, wrong field, too vague, unsupported by the cited quote,
or upstream OCR/chunking failure) and with explicit \emph{missing} records
where the paper states something the extractor reported as absent. Facts
carrying a machine citation flag were reviewed first, so the labeled subset is
not a random sample. The rates in Table~\ref{tab:gold-set} describe the
labeled subset.

\begin{table}[t]
\caption{Human verdicts on the extraction gold set, by the status the
extractor assigned. ``Citation defect'' is the reviewer's OCR/chunking
verdict: the value is right but the quotation is not a sound citation.}
\label{tab:gold-set}
\centering
\footnotesize
\begin{tabular}{lrrrr}
\hline
Extractor status & $n$ & Correct & Citation defect & Other \\
\hline
Supported & 268 & 186 & 27 & 55 \\
Ambiguous & 45 & 18 & 14 & 13 \\
Not found & 19 & 18 & 0 & 1 \\
Unsupported & 2 & 2 & 0 & 0 \\
\hline
\end{tabular}
\end{table}

\paragraph{Extraction: verdicts}
Three results describe the extractor. First, its own status is informative
but not decisive. Of the 268 facts it marked supported, 186 (69\%) were
correct with a sound citation. Of the 45 it had downgraded to ambiguous, most
for a missing or unlocatable citation, 18 were right values the model had
failed to cite. Of the 19 it reported as not found, 18 were confirmed absent
from the reading surface. Second, the citation is a distinct failure mode
from the value. Twenty-seven supported facts (10\%) received the OCR/chunking
verdict: the value is right while the quotation is cited to the wrong chunk of
the same paper, duplicated, stitched from separate fragments, paraphrased, or
corrupted by extraction metadata. Counting those, the value is correct for 213
of the 268 supported facts (79\%); wrong field, too vague, and unsupported
account for the other 55. Third, a second-pass judge, the same model asked
independently whether the stored passage supports the stored value, flags 32
of the 69 value errors and 30 of the 224 correct facts, so it is a triage aid
and not a verdict.

\paragraph{Extraction: mechanical checks against the verdicts}
For these reasons the catalog does not store the model's transcription of its
evidence. A quotation is located as one compact passage of a chunk and stored
as the source text with its character offsets, and a quotation that the model
joined with an ellipsis is stored as its separate passages. A quotation that
cannot be located in this way stays flagged for review. Of the 41 labeled
facts with a citation defect, 16 now carry located source text (the locator
repaired the citation after the verdict was given), 19 carry a quotation that
cannot be located as a passage of the paper, and six carry none. The
mechanical check flags 36 of the labeled facts, 30 of which (83\%) the
reviewers had judged negative; it catches 19 of the 41 citation defects.

These results set the wording of the claim. A located quotation is source text
with its position in the paper, which shows that the text is there. Whether it
supports the value remains a human judgment. No extracted fact or proposed edge is treated as more than reviewable
candidate evidence before that review.

\paragraph{Consequence of the review: a revised extraction prompt}
The verdicts and the reviewers' notes were turned into a revised extraction
prompt, run on the same 33 papers, the same chunks, and the same model. The
revision requires a short exact quotation for every fact and marks a fact
without one as ambiguous rather than supported, asks for one fact per stated
item (three instruments are three facts), keeps a stated temperature range
whole, and adds sample form, material alias, and a per-fact subject so that a
value no longer carries the material it belongs to in its text.
Table~\ref{tab:extraction-revision} separates the mechanical changes across
the full extraction from the human verdicts on the reviewed subsets.
Matching facts by value and citation overlap carries 141 earlier verdicts
onto the revised extraction, 47 on identical values.

\begin{table}[t]
\caption{Original and revised extraction on the same 33 papers, chunks, and
model. The original review prioritized flagged facts; the revised sample was
drawn randomly from records without transferred labels. Because the sampling
procedures differ, the percentages are not directly comparable estimates of
overall extraction accuracy.}
\label{tab:extraction-revision}
\centering
\footnotesize
\setlength{\tabcolsep}{3pt}
\begin{tabular}{p{0.44\columnwidth}rr}
\hline
Measure & Original & Revised \\
\hline
\multicolumn{3}{l}{Full extraction} \\
Facts & 492 & 868 \\
Marked supported & 401 & 841 \\
Located quotations & 503/545 & 852/860 \\
Downgraded for missing citation & 40 & 0 \\
\hline
\multicolumn{3}{l}{Human review of supported facts} \\
Facts reviewed & 268 & 59 \\
Correct with sound citation & 186 (69\%) & 56 (95\%) \\
Correct value, defective citation & 27 & 1 \\
Other verdicts & 55 & 2 \\
\hline
\end{tabular}
\end{table}

Following revision, one coauthor reviewed a new random sample of 60 extraction
records without transferred labels, using the same five verdicts. Excluding
one machine-stamped duplicate left 59 records, all marked supported by the
extractor. Of these newly reviewed records, 56 (95\%) were judged correct with
a sound citation. One has a correct value but an OCR defect in the quoted source
(a decimal point rendered as a colon). Two have the wrong field: an
abbreviation stored as a composition and an applied strain stored as a
sample form. No sampled value was judged wrong, too vague, or unsupported.
This audit measures extraction quality in the intended collection of published,
BT-7-linked papers.
The case studies use the original extraction.

\paragraph{Paper-to-experiment linkage}
The linker of Section~\ref{sec:workflows} ranks candidate experiments for a
paper. A reviewer judged the five highest-ranked
candidates for each of 48 BT-7 papers. Twelve papers were matched to their
experiment, twelve could not be decided from the available evidence, and for
twenty-four none of the five candidates was the paper's experiment. Where the
match is a whole experiment, the linker had ranked it first for four of the
five papers and fifth for one; the other matches are to scan groups within an
experiment. A human verdict overrides the score of its candidate; a candidate
without a verdict carries an uncalibrated score. These verdicts are the
calibration signal for the linkage priors.

\section{Discussion and outlook}
\label{sec:discussion}

The two cases show why evidence must remain inspectable as a task moves between tools. In the SrMn$_2$Sb$_2$ analysis, analog literature informed a human-selected model family, historical records established the experimental context, and resolution-aware fitting yielded quantitative exchange estimates under explicit model and resolution assumptions. In the live BT-7 case, the same record structure connected an acquisition policy to achieved positions, measured counts, and an independent reference experiment. The contribution is this continuity: a result retains the evidence and decisions needed to evaluate it after the workflow has moved on.

\subsection{Transfer to other instruments}

Other characterization methods also combine heterogeneous logs, incomplete metadata, external modeling codes, and human interpretation. The explicit representation of missing inputs, rejected candidates, calibration state, review decisions, and source lineage provides a basis for extending the methodology demonstrated on BT-7 to other facilities.

Recording these states as data applies FAIR principles \citep{wilkinson2016fair} and provenance vocabularies such as PROV-O \citep{provo2013} to decisions made during analysis and measurement. A subsequent user can inspect both the result and the conditions under which it was accepted. This complements curated materials resources such as the Materials Project \citep{jain2013materialsproject} by connecting published knowledge to local experimental evidence and calculations.

Existing reader adapters already support data formats from multiple TAS facilities. Extending complete workflows to an instrument such as HB-1A at the High Flux Isotope Reactor, Oak Ridge National Laboratory, would build on those readers and the shared contracts, review store, and workflow definitions. The next steps are to validate instrument-specific axes, scan families, configuration, and calibration conventions against local records. More complete metadata in future data files would also make retrospective analysis easier: information absent from an archive may remain available to the original experimenters or obtainable from them.

\subsection{Facility workflows}

The system is in service at a user facility: its planning pages are public, its extraction and linkage layers are assessed against human verdicts, and it has been used for fine alignment and adaptive measurement on BT-7. The cases demonstrate a methodology for connecting published evidence, experimental records, and scientific tools in facility workflows.

Executable reduction extends that connection to the calculations themselves. Recorded dependencies determine which results must be reconsidered when a background, correction, or processing parameter changes, while preserving the unchanged parts of the analysis. Extending this treatment to additional fitting and detector-reconstruction workflows would support consistent revision of an analysis from raw scans through its final interpretation.

\subsection{Conclusion}

TAS Co-pilot connects literature, experimental records, scientific calculations, and measurement through shared contracts and provenance. The retrospective and live cases demonstrate how those records support analysis, expose execution and inference failures, and preserve the limits of a result. This makes the basis of a scientific decision available for inspection and reuse, while keeping model-generated proposals subject to validation and human review.

\section*{Author Contributions}
W.D.R. conceived the project and designed the architecture. C.J. and S.K.
developed the paper-to-experiment linkage and the review of extracted paper
facts: C.J. reviewed the linkage between the BT-7 publication corpus and the
experiment catalog and carried out the SrMn$_2$Sb$_2$ spin-wave analysis, and
S.K. produced the human-labeled gold set used in the evaluation. L.P. and
J.W.L. collected the SrMn$_2$Sb$_2$ data of experiment 24858. J.W.L.
reviewed the retrospective analysis. A.P. developed
the MCP tool gateway, the integration with the NIST-hosted language-model
endpoint, and the graph-based retrieval over the evidence graph. W.D.R.
performed the live BT-7 case. C.J., A.P., S.K., J.W.L., and W.D.R. contributed
to the code, analyzed the results, and wrote the manuscript.

\section*{Acknowledgments}
The authors thank Johnpierre Paglione of the University of Maryland.
The authors acknowledge the Texas Advanced Computing Center (TACC) at The
University of Texas at Austin for resources made available to NIST under
contract number 1333ND25PNB180410 that have contributed to the research
results reported within this paper. URL: \url{http://www.tacc.utexas.edu}.
Support for Carina Jacobson, Aditya Purohit, and Srikar Kovvali was provided
by the Center for High Resolution Neutron Scattering (CHRNS), a partnership
between the National Institute of Standards and Technology and the National
Science Foundation under Agreement No.~DMR-2010792.
Anthropic Claude and OpenAI Codex were used for coding assistance, including
the development of figure-generation scripts, and for drafting and substantive
editing of the manuscript. The authors directed the use of these tools,
reviewed and checked the manuscript, code, and results, and take responsibility
for their accuracy and integrity. The research data supplied to the foundation
models were publicly available online.
Certain commercial products are identified in this paper in order to specify
the computational procedures adequately. Such identification is not intended
to imply recommendation or endorsement by the National Institute of Standards
and Technology, nor is it intended to imply that the products identified are
necessarily the best available for the purpose.

\section*{Conflicts of Interest}
The authors declare no conflicts of interest.

\section*{Data Availability Statement}
The raw BT-7 data that support the findings of this study are openly
available in the NIST Center for Neutron Research data repository:
experiment \texttt{24858} (SrMn$_2$Sb$_2$, 2018) at
\url{https://ncnr.nist.gov/pub/ncnrdata/bt7/201804/24858/} and
\url{https://ncnr.nist.gov/pub/ncnrdata/bt7/201806/24858/}, the live
La$_{0.7}$Sr$_{0.3}$MnO$_3$ case at
\url{https://ncnr.nist.gov/pub/ncnrdata/bt7/202608/14/}, and its 2017
reference measurement at
\url{https://ncnr.nist.gov/pub/ncnrdata/bt7/201706/23816/}.
The TASCopilot source code and reproducibility materials will be publicly
available on GitHub upon journal publication.

Copyrighted publisher PDFs, extracted paper text, and the
derived corpus catalogs are not redistributed.

\section*{Supporting Information}
Supporting Information accompanies this article.

\bibliography{references}

\end{document}